\documentclass{aa} 
\usepackage{graphicx,natbib} 
\usepackage{txfonts} 
\usepackage{longtable}

\bibpunct{(}{)}{;}{a}{}{,} 
 
\def\h2o{H$_2$O}
 
\begin{document} 
 
\title{Magnetic fields around evolved stars: further observations of \h2o 
maser polarization} 
 
\author{Leal-Ferreira, M. L.\inst{1} \and Vlemmings, W. H. T.\inst{2} \and Kemball, A.\inst{3,4} \and Amiri, N.\inst{5}
}

\institute{Argelander-Institut f\"ur Astronomie, Universit\"at Bonn, 
Auf dem H\"ugel 71, 53121 Bonn, Germany\\ 
\email{ferreira@astro.uni-bonn.de} 
\and Department of Earth and Space Sciences, Chalmers University of Technology, Onsala Space Observatory, SE-439 92 Onsala, Sweden
\and Department of Astronomy, University of Illinois at Urbana-Champaign, 1002 West Green Street, Urbana, IL 61801, USA 
\and National Center for Supercomputing Applications, University of Illinois at Urbana-Champaign, 605 East Springfield Avenue, Champaign, IL 61820, USA 
\and Center for Astrophysics and Space Astronomy, Department of Astrophysical and Planetary Sciences, University of Colorado, 389 UCB, Boulder, CO 80309-0389, USA
}
\date{received , accepted} 
\authorrunning{Leal-Ferreira et al.} 
\titlerunning{Magnetic fields around evolved stars: \h2o maser polarization} 

\abstract 
{A low- or intermediate-mass star is believed to maintain a spherical shape 
throughout the evolution from the main sequence to the Asymptotic Giant Branch 
(AGB) phase. However, many post-AGB objects and planetary nebulae exhibit 
non-spherical symmetry. Several candidates have been suggested as factors 
that can play a role in this change of morphology, but the problem is still 
not well understood. Magnetic fields are one of these possible agents.}
{We aim to detect the magnetic field and infer its properties around four 
AGB stars using \h2o maser observations. The sample we observed consists of 
the following sources: the semi-regular variable RT~Vir, and the Mira 
variables AP~Lyn, IK~Tau, and IRC$+$60370.}
{We observed the 6$_{1,6}-$5$_{2,3}$ \h2o maser rotational transition in 
full-polarization mode to determine its linear and circular polarization. 
Based on the Zeeman effect, one can infer the properties of the 
magnetic field from the maser polarization analysis.}
{We detected a total of 238 maser features in three of the four observed 
sources. No masers were found toward AP~Lyn. The observed masers are all 
located between 2.4 and 53.0~AU from the stars. Linear and circular 
polarization was found in 18 and 11 maser features, respectively.}
{We more than doubled the number of AGB stars in which a magnetic field 
has been detected from \h2o maser polarization. Our results confirm the 
presence of fields around IK~Tau, RT~Vir, and IRC$+$60370. The strength 
of the field along the line of sight is found to be between 47 and 
331~mG in the \h2o maser region. Extrapolating this result to the surface 
of the stars, assuming a toroidal field ($\propto$~r$^{-1}$), we 
find magnetic fields of 0.3$-$6.9~G on the stellar surfaces. 
If, instead of a toroidal field, we assume a poloidal field 
($\propto$~r$^{-2}$), then the extrapolated magnetic field strength on the 
stellar surfaces are in the range between 2.2 and $\sim$115~G. Finally, if 
a dipole field ($\propto$~r$^{-3}$) is assumed, the field strength on the 
surface of the star is found to be between 15.8 and $\sim$1945~G. The magnetic 
energy of our sources is higher than the thermal and kinetic energy in the 
\h2o maser region of this class of objects. This leads us to conclude that, 
indeed, magnetic fields probably play an important role in shaping the 
outflows of evolved stars.}

\keywords{masers, polarization, magnetic field, Stars: AGB and post-AGB} 
 
\maketitle 
 
\section{Introduction} 

Low- and intermediate-mass stars (0.8$-$8~M$_\odot$) are believed to maintain 
their sphericity until the asymptotic giant branch (AGB) phase. Even though 
some AGB stars are slightly elliptical \citep[e.g.,][]{reid07,castro10}, 
many planetary nebulae (PNe) do not present any spherical symmetry. How an 
almost-spherical AGB star gives rise to a non-spherical PN is still an 
open question. A companion to the star (binary system or a massive planet), 
disk interaction, the influence of magnetic fields, or a combination of 
these agents are candidates to explain this phenomenon \citep[][~and 
references therein]{balick02,frank07,nordhaus07}.

Magneto-hydrodynamic (MHD) simulations show that the magnetic field can be 
an important agent in shaping post-AGBs and PNe \cite[e.g.,][]{garcia99,
garcia05,garcia08,dennis09}. Moreover, recent observations support the 
presence of magnetic fields around AGB and post-AGB stars 
\cite[e.g.,][]{amiri11,perez11,lf12,vlemmings12}. However, the sample of low 
and intermediate mass evolved stars around which magnetic fields have been 
measured is still small. So far, detections of magnetic field 
from \h2o maser polarization were reported around two AGB stars only; U~Her 
and U~Ori \citep{vlemmings02,vlemmings05}. Also, the morphology and strength 
of the magnetic field as a function of radial distance throughout the 
circumstellar envelope is still unclear. Observations of different magnetic 
field tracers are needed to constrain the field dependence on the radial 
distance from the star and, therefore, improve future MHD simulations.

Different maser species can provide information about different regions 
around these objects. While SiO masers are expected to be found within 
the extended atmosphere of the star (between the photosphere and the 
dust formation zone), OH masers are detected much further out 
($\sim$65$-$650~AU). The \h2o masers emit at an intermediate distance 
to the star, between the SiO and OH maser regions. The distance of the 
\h2o masers from the star is expected to lie within a few to less than a 
hundred AU \cite[e.g.,][]{cohen87,bowers89,elitzur92}.

The present work aims to enlarge the number of magnetic field detections 
around low- and intermediate-mass evolved stars. We imaged five sources 
of this class using very-long-baseline interferometry (VLBI), in 
full-polarization mode, with the goal of detecting \h2o masers around 
them. As a result of Zeeman splitting \citep{zeeman97}, we can measure 
the magnetic field signature on maser lines by investigating the 
polarized emission of the masers \citep[e.g.,][]{vlemmings01,vlemmings06}. 

Our sample is composed of the pre-PN OH231.8$+$4.2, the semi-regular variable 
RT~Vir, and the Mira variables AP~Lyn, IK~Tau, and IRC$+$60370. We presented 
the results of OH231.8$+$4.2 in \citet{lf12}. The analysis of the four 
remaining sources is presented in the present paper. Single-dish SiO maser 
observations in full-polarization mode have been previously reported by 
\cite{herpin06} for RT~Vir, AP~Lyn, and IK~Tau. Their results show a magnetic 
field of 0~$\leq$~$B_{||}$[G]~$\leq$~5.6 in RT~Vir, 
0.9~$\leq$~$B_{||}$[G]~$\leq$~5.6 in AP~Lyn, and 
1.9~$\leq$~$B_{||}$[G]~$\leq$~6.0 in IK~Tau. The AGB star RT~Vir also shows 
strong circular polarization in single dish OH maser observations, indicating 
a strong global magnetic field \citep{szymczak01}. We did not find any 
literature reports concerning the magnetic field for IRC$+$60370 in the 
SiO maser region, nor for AP~Lyn, IK~Tau, and IRC$+$60370 in the OH maser 
region. 

This paper is structured as follows: in Sect.~2, we describe the 
observations, data reduction, and calibration; in Sect.~3, we present the 
results; in Sect.~4, we discuss the results and, in Sect.~5, we conclude 
the analysis.

\section{Observations and data reduction} 
We used the NRAO\footnote{The National Radio Astronomy Observatory (NRAO) 
is a facility of the National Science Foundation operated under cooperative 
agreement by Associated Universities, Inc.} Very Long Baseline Array (VLBA) 
to observe the \h2o 6$_{1,6}-$5$_{2,3}$ rotational maser transition at a rest 
frequency toward 22.235081~GHz of the stars in our sample. In each observing 
run, we used two baseband filters and performed separate lower (Low) and 
higher (High) resolution correlation passes. The first was performed 
in full-polarization mode and the second in dual-polarization mode. We 
show the characteristics of the Low and High correlation passes in 
Table~\ref{observres} and the individual observation details of each source 
in Table~\ref{observsource}. 

\begin{table}
  \begin {center}
    \caption{Low- and high-resolution correlation passes}
    \begin{tabular}{@{}ccccc@{}}
      \hline
Label & N$_{chans}$ & BW  & $\triangle v$   & PolMode\\
      &       & (MHz) & (km/s) & \\
      \hline
Low    & 128   & 1.0   & 0.104  & Full (LL,RR,LR,RL)\\
High    & 512   & 1.0   & 0.026  & Dual (RR,LL)\\
      \hline
    \end{tabular}
    \label{observres}
    \end{center}
{Correlation parameters for the low- and high-resolution correlation passes. 
Description of Cols. 1 to 5: The label of the observed data $-$ low- (Low) 
and high- (High) resolution $-$ (Label), the number of channels (N$_{chans}$), 
the bandwidth (BW), the channel width ($\triangle v$), and the polarization 
mode (PolMode).}
\end{table}

\begin{table*}
  \begin{center}
    \caption{Source observation details}
    \begin{tabular}{@{}ccccccccc@{}}
      \hline
Code  & Source      & Class               & V$_{lsr}$(IF1)&V$_{lsr}$(IF2)& Beam                   &RA$_0$              & Dec$_0$             & Date \\
      &             &                     & (km/s)       & (km/s)      & (mas)                  & (J2000)            & (J2000)             &(mm/dd/yy)\\
      \hline
BV067A*&OH231.8$+$4.2& pre-Planetary Nebula&$+$44.0      & $+$26.0     &1.7$\times$0.9          &07$^h$42$^m$16.93$^s$& --14$^\circ$42'50''.2& 03/01/09\\
BV067B & AP~Lyn      & Mira variable       & --19.5      & --32.5      & --                     &06$^h$34$^m$34.88$^s$&$+$60$^\circ$56'33''.2& 03/15/09\\
BV067C & IK~Tau      & Mira variable       &$+$42.5      & $+$29.5     &1.2$\times$0.5          &03$^h$53$^m$28.84$^s$&$+$11$^\circ$24'22''.6& 02/20/09\\
BV067D & RT~Vir      &Semi-regular variable&$+$25.5      & $+$12.5     &1.2$\times$0.9&13$^h$02$^m$37.98$^s$&$+$05$^\circ$11'08''.4& 03/15/09\\
BV067E & IRC$+$60370 & Mira variable       & --44.5      & --57.5      &0.8$\times$0.5          &22$^h$49$^m$58.88$^s$&$+$60$^\circ$17'56''.7& 03/05/09\\
\hline
\multicolumn{7}{l}{*Presented in \citet{lf12}}\\
      \hline
    \end{tabular}
    \label{observsource}
    \end{center}
{From left to right: The project code (Code), the name of 
the source (Source), the nature of the source (Class), the velocity center 
position of each of the 2 filters (v$_{lsr}$), the PSF beam size (Beam), the 
center coordinates of the observations (RA$_0$ and Dec$_0$), and the starting 
observation date (Date).}
\end{table*}

We observed different calibrators for each target. Each calibrator was
observed during the same run as its corresponding target. For the
calibration of RT~Vir, we used 3C84 (bandpass, delay, polarization 
leakage, and amplitude).  To calibrate IK~Tau, we used J0238$+$16 
(bandpass, delay, and amplitude) and 3C84 (polarization 
leakage). To calibrate IRC$+$60370, we used BLLAC (bandpass, delay, 
polarization leakage, polarization absolute angle, and amplitude). 
Unfortunately, no good absolute polarization angle calibrator were 
available for RT~Vir and IK~Tau, making it impossible to determine 
the absolute direction of the linear polarization vectors (also 
referred to as electric vector position angle; EVPA). However,
the relative EVPA angles for individual polarized components within
RT~Vir are still correct (no linear polarization was detected for 
IK~Tau). To determine the absolute EVPA of IRC$+$60370, we created 
a map of BLLAC and compared the direction of the measured EVPA with 
that reported in the VLA/VLBA polarization calibration
database\footnote{http://www.vla.nrao.edu/astro/calib/polar/2009/K$\_$band$\_$2009.shtml}.
Our IRC$+$60370 observation was carried out between the calibration
observations of February 21 and March 19, 2009 in that
database, where the polarization angle of BLLAC changed from 25.7$^\circ$ to
26.0$^\circ$. We thus adopted a reference angle of 25.8$^\circ$ to obtain 
the absolute EVPA.

After an initial analysis of the raw data, we did not detect any maser emission 
around AP~Lyn and so did not proceed with further calibration of this data 
set. For the other three targets, we used the Astronomical Image Processing 
Software Package (AIPS) and followed the data reduction procedure documented 
by \citet{kemball95} to perform all the necessary calibration
steps. This included using the AIPS task {\it SPCAL} to determine
polarization leakage parameters using a strong maser feature.

After the data were properly calibrated, we used the low-resolution data to 
create the image cubes for the Stokes parameters $I$, $Q$, $U$, and $V$. The 
$Q$ and $U$ cubes were used to generate the linear polarization intensity 
($P=\sqrt[]{Q^2+U^2}$) cubes and the EVPA cubes. The noise level 
measured on the emission-free channels of the low-resolution data cubes is 
between $\sim$2~mJy and $\sim$6~mJy. The high-resolution data were used to 
create the data cubes of the Stokes parameters $I$ and $V$, from which the 
circular polarization could be inferred. The noise level measured from the 
emission-free channels of the high-resolution data cubes is between 
$\sim$5~mJy and $\sim$11~mJy.

The detection of the maser spots was done by using the program {\it maser 
finder}, as described by \citet{surcis11}. We defined a maser feature 
to be successfully detected when maser spots located at similar spatial 
positions (within the beam size) survive the signal-to-noise ratio cutoff 
we adopted (8~$\sigma$) in at least three consecutive channels. The position 
of the maser feature was taken to be the position of the maser spot in the 
channel with the peak emission of the feature \cite[see e.g.,][]{richards11}.

\section{Results} 

We found 85 maser features around IK~Tau, 91 toward RT~Vir, and 62 around 
IRC$+$60370. The maser identification and properties are shown in Table 
\ref{results}. In Fig.~\ref{masermaps} we show the spatial distribution of 
the maser components (depicted as circles). The size of the circles is 
proportional to the maser flux densities, and they are colored according to 
velocity. The black cross indicates the stellar position determined in 
Sect.~\ref{starpos}.

Positive linear polarization detection is reported when successfully found 
in at least two consecutive channels. The linear polarization percentage 
($P_L$) quoted in Table \ref{results} is the $P_L$ measured in the 
brightest channel of the feature. The $P_L$ error is given by the rms of 
the $P$ spectrum on the feature spatial position, scaled by the intensity 
peak. The $EVPA$ error was determined using the expression $\sigma_{EVPA} = 
0.5~ \sigma_{P}/P \times 180^\circ/\pi$ \citep{wardle74}. The linear 
polarization results are enumerated in Cols.~8 ($P_L$) and 9 ($EVPA$) of 
Table \ref{results}. In Fig. \ref{masermaps}, the black vectors show 
the $EVPA$ of the features in which linear polarization is present. The 
length of the vectors is proportional to the polarization percentage.

To measure the circular polarization, we used the $I$ and $V$ spectra to 
perform the Zeeman analysis described by \cite{vlemmings02}. In this approach, 
the fraction of circular polarization, $P_V$, is given by
\begin{eqnarray}
P_V &=& (V_{max}~-~V_{min})/I_{max} \nonumber \\
    &=& 2~\times A_{F-F'}~\times B_{||}[Gauss] /\Delta v_L[km/s],
\label{eq1}
\end{eqnarray}

\noindent where $V_{max}$ and $V_{min}$ are the maximum and minimum of the model 
fitted to the $V$ spectrum, and $I_{max}$ is the peak flux of the emission. 
The variable $A_{F-F'}$ is the Zeeman splitting coefficient. Its exact 
value depends on the relative contribution of each hyperfine component of the 
\h2o 6$_{1,6}-$5$_{2,3}$ rotational maser transition. We adopted the value 
$A_{F-F'}$~=~0.018, which is the typical value found by \cite{vlemmings02}. 
The projected magnetic field strength along the line of sight is given by 
$B_{||}$ and $\Delta v_L$ is the full-width half-maximum of the $I$ spectrum. 
Although the non-LTE analysis in \cite{vlemmings02} has shown that the 
circular polarization spectra are not necessarily strictly proportional 
to $dI/d \nu$, using $A_{F-F'}$, determined by a non-LTE fit, introduces a 
fractional error of less than $\sim$20$\%$ when using Eq.~1. We report 
circular polarization detection when the magnetic field strength given by 
the model fit is $\geq$ 3$\sigma$. The reported errors are based on the 
single channel rms using Eq.~\ref{eq1} \cite[see][ Sect.~3.3, for further 
discussion]{lf12}. We list the $P_V$ and $B_{||}$ results in Cols.~10 and 11 
of Table~\ref{results}, where the positive sign on $B_{||}$ indicates that 
the direction of the magnetic field along the line of sight is away from 
the observer, while the negative sign corresponds to a direction towards 
the observer. In Figs.~\ref{circpol} and \ref{rt90}, we present the $I$ 
and $V$ spectra and the model fit of $V$ spectra for those features in which 
we detect circular polarization.

\subsection{IK~Tau}

\begin{figure*}
 \centering
 \begin{tabular}{c}
   \includegraphics[width = 139 mm]{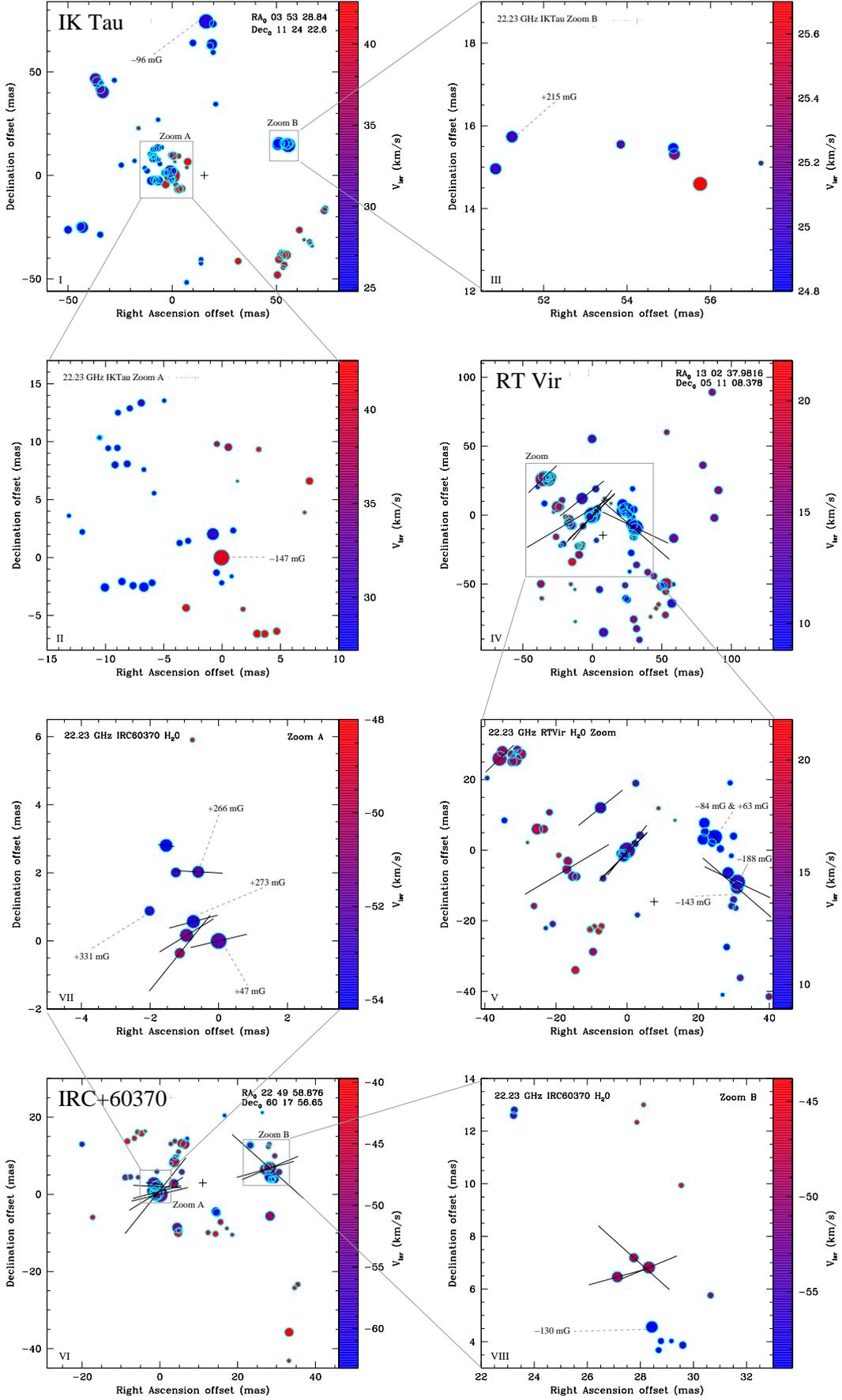} \\
 \end{tabular}
 \caption[]{Maps of the maser features detected toward RT~Vir, IK~Tau, and 
IRC$+$60370. The size of the circles is proportional to the maser flux 
densities, and their colors show the velocity scale. The black crosses 
indicate the stellar positions (see Sect.~\ref{starpos}). The black lines 
indicate the EVPA (for RT~Vir, they could not be calibrated in terms of 
absolute EVPA), and their lengths are proportional to the fractional linear 
polarization. The magnetic field strength along the line of sight are also 
shown for the features in which we detected circular polarization. The x-axis 
is the projected offset on the plane of the sky in the direction of right 
ascension. The y-axis is the declination offset. The offsets are with 
respect to the reference maser.}
   \label{masermaps}
 \end{figure*}

\begin{figure*}
\centering
\begin{tabular}{lll}
  \includegraphics[width = 59 mm]{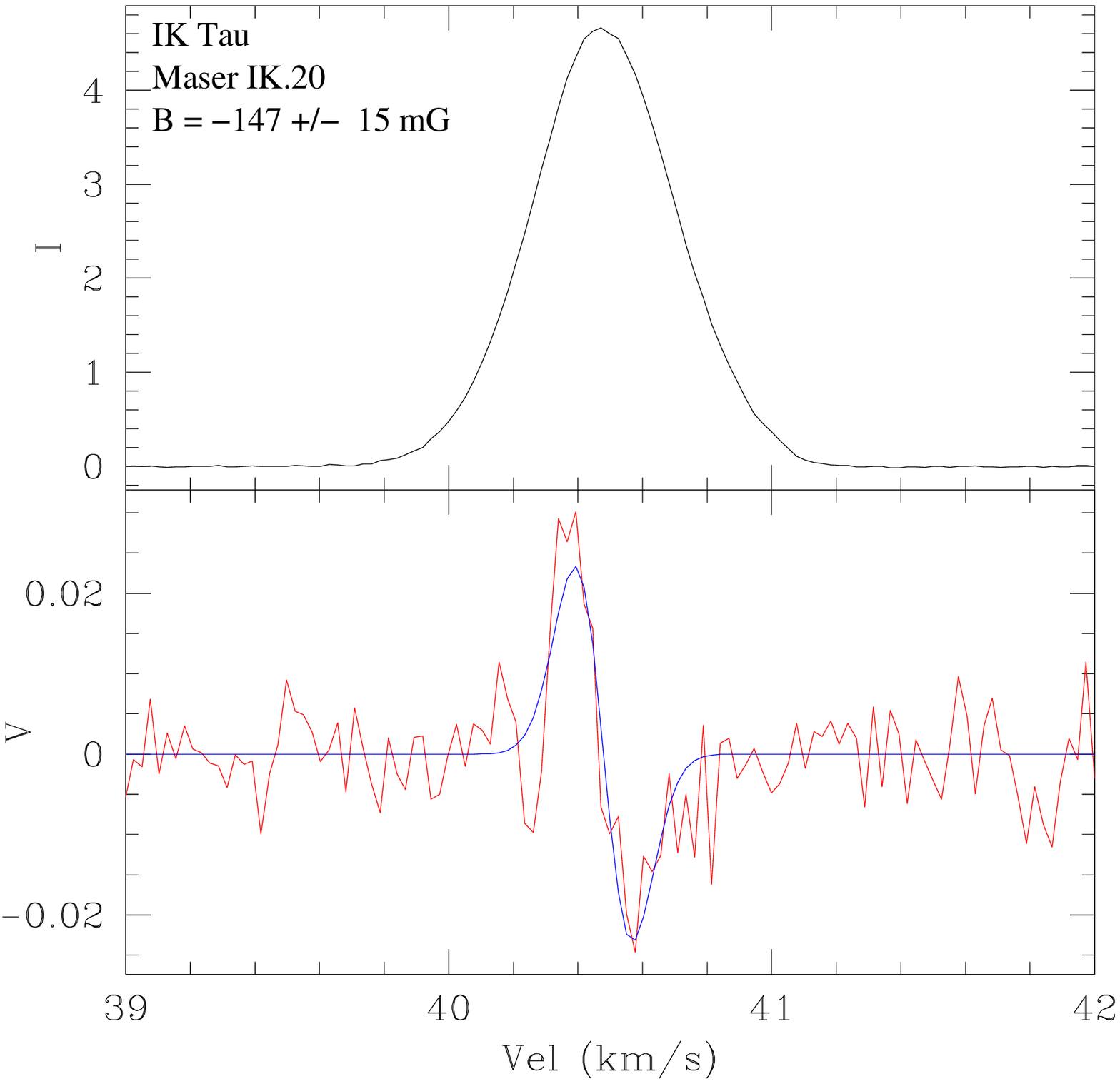} &
  \includegraphics[width = 59 mm]{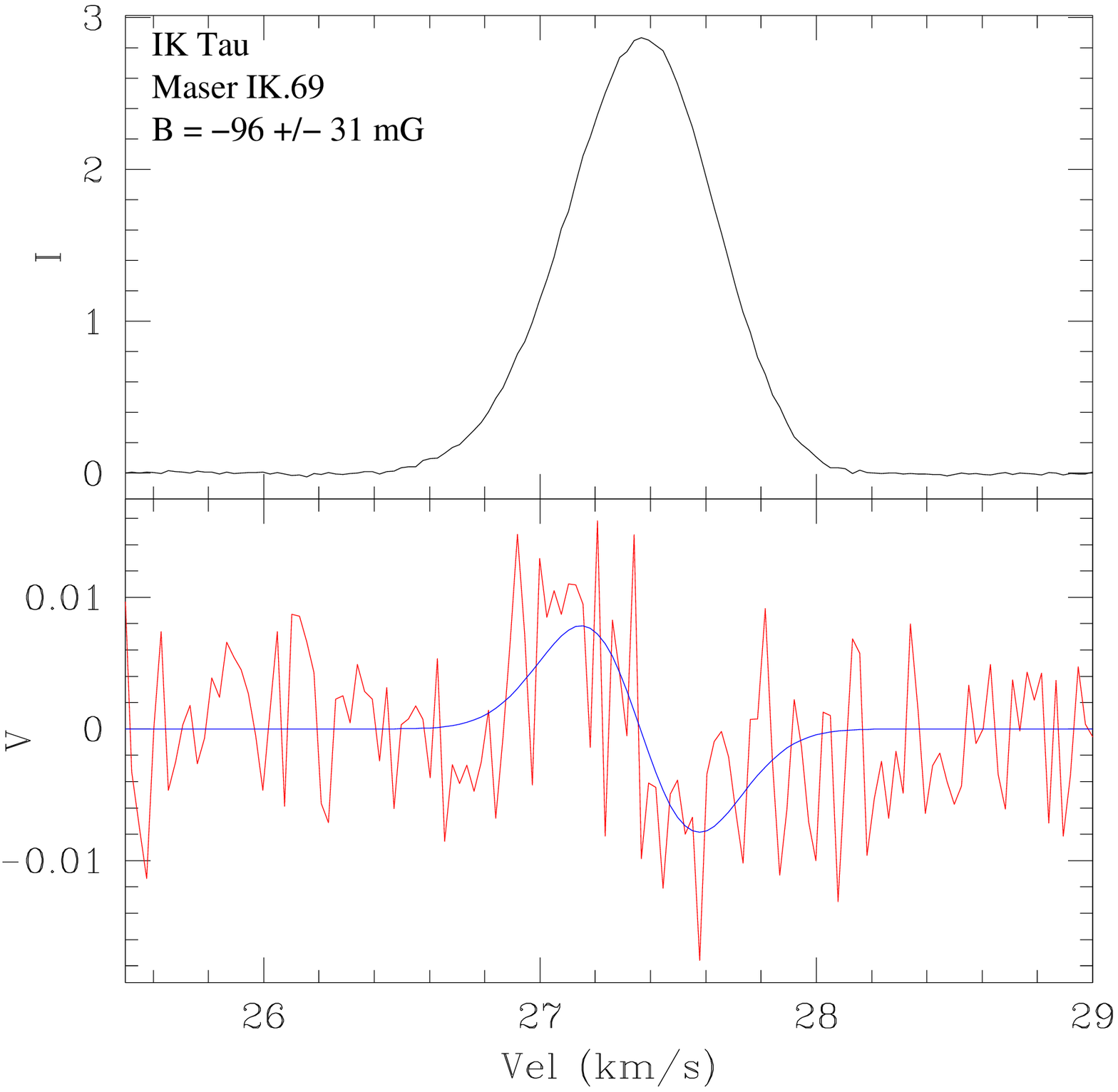} &
  \includegraphics[width = 59 mm]{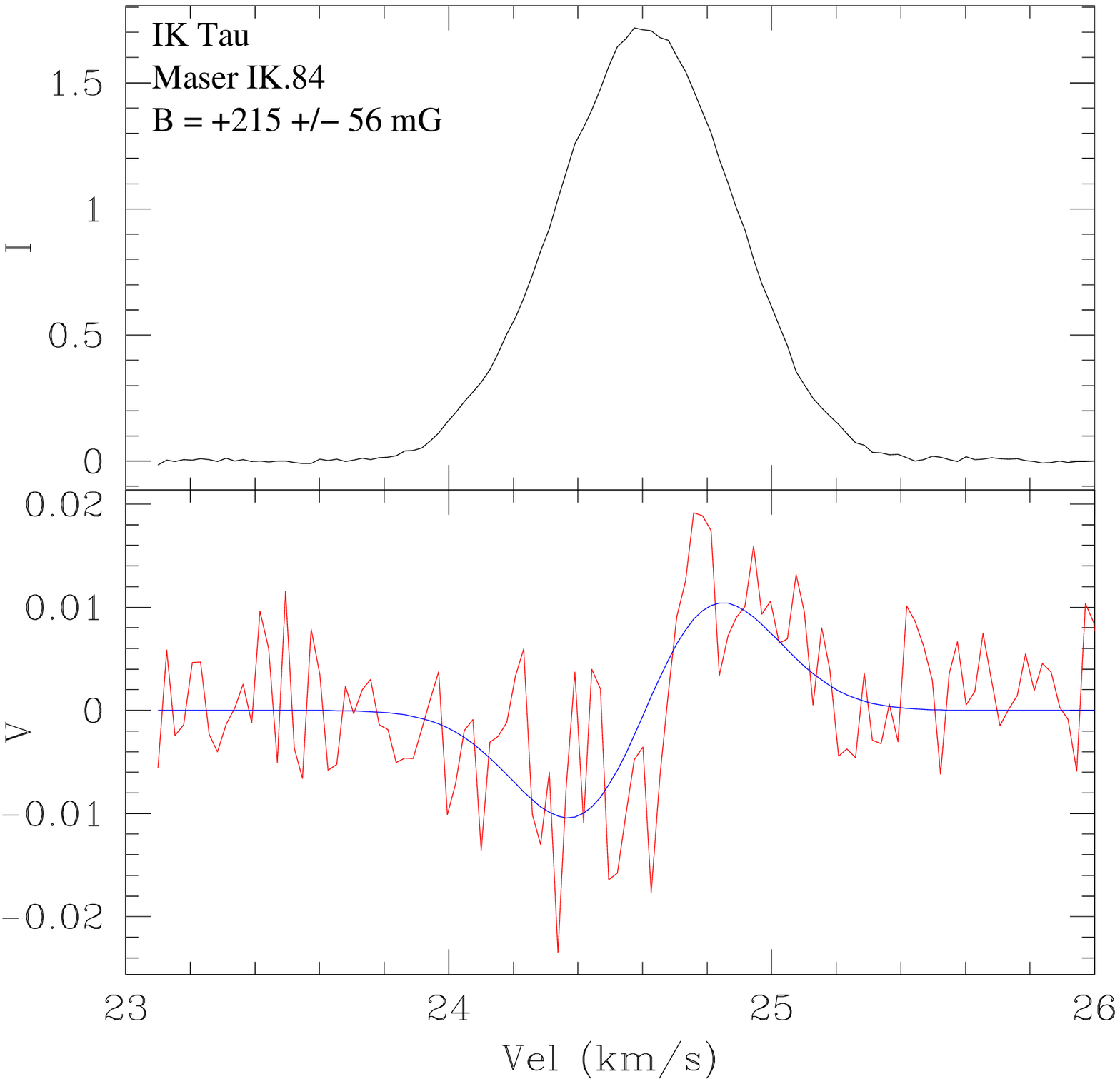} \\
  \includegraphics[width = 59 mm]{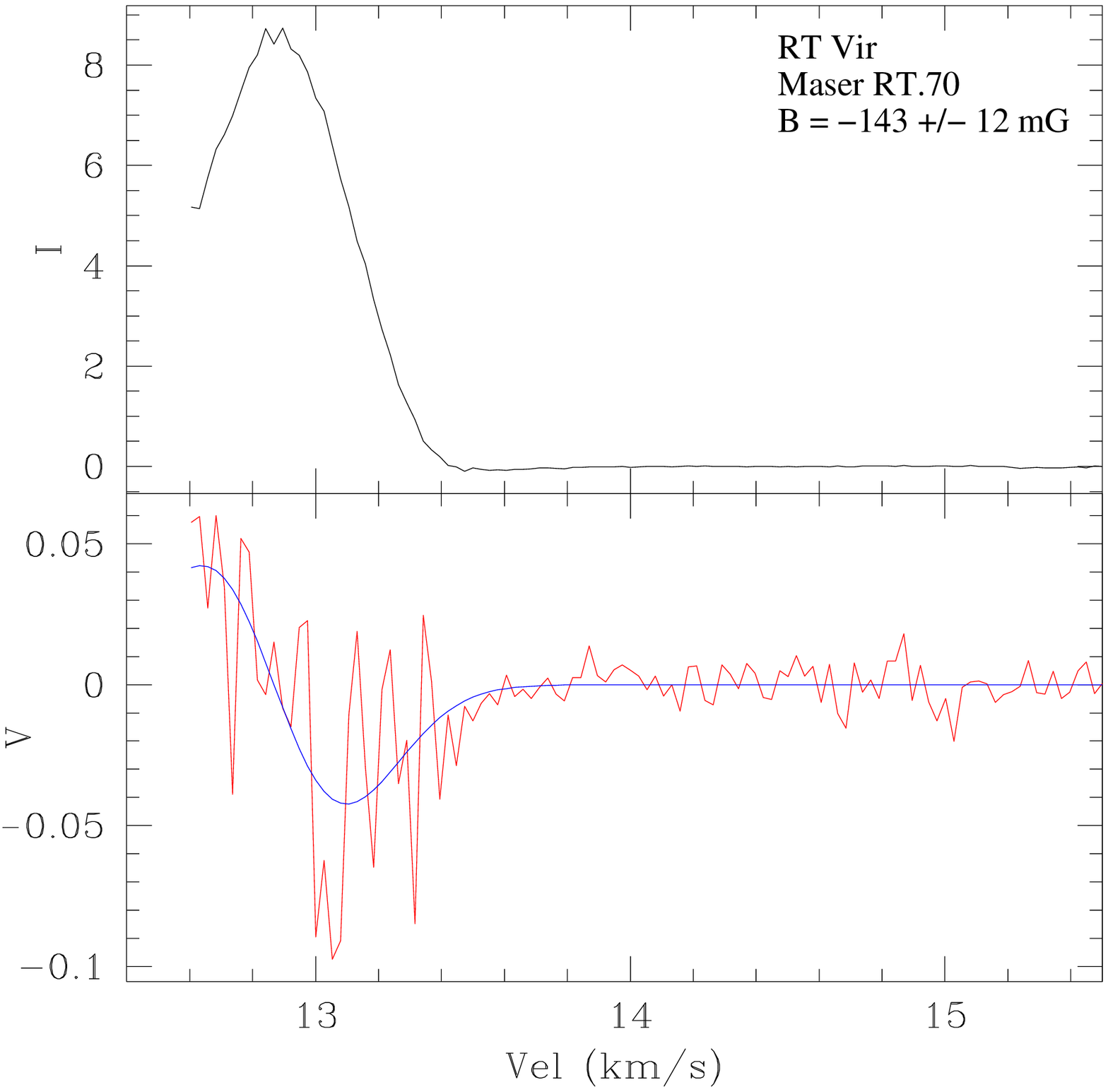} &
  \includegraphics[width = 59 mm]{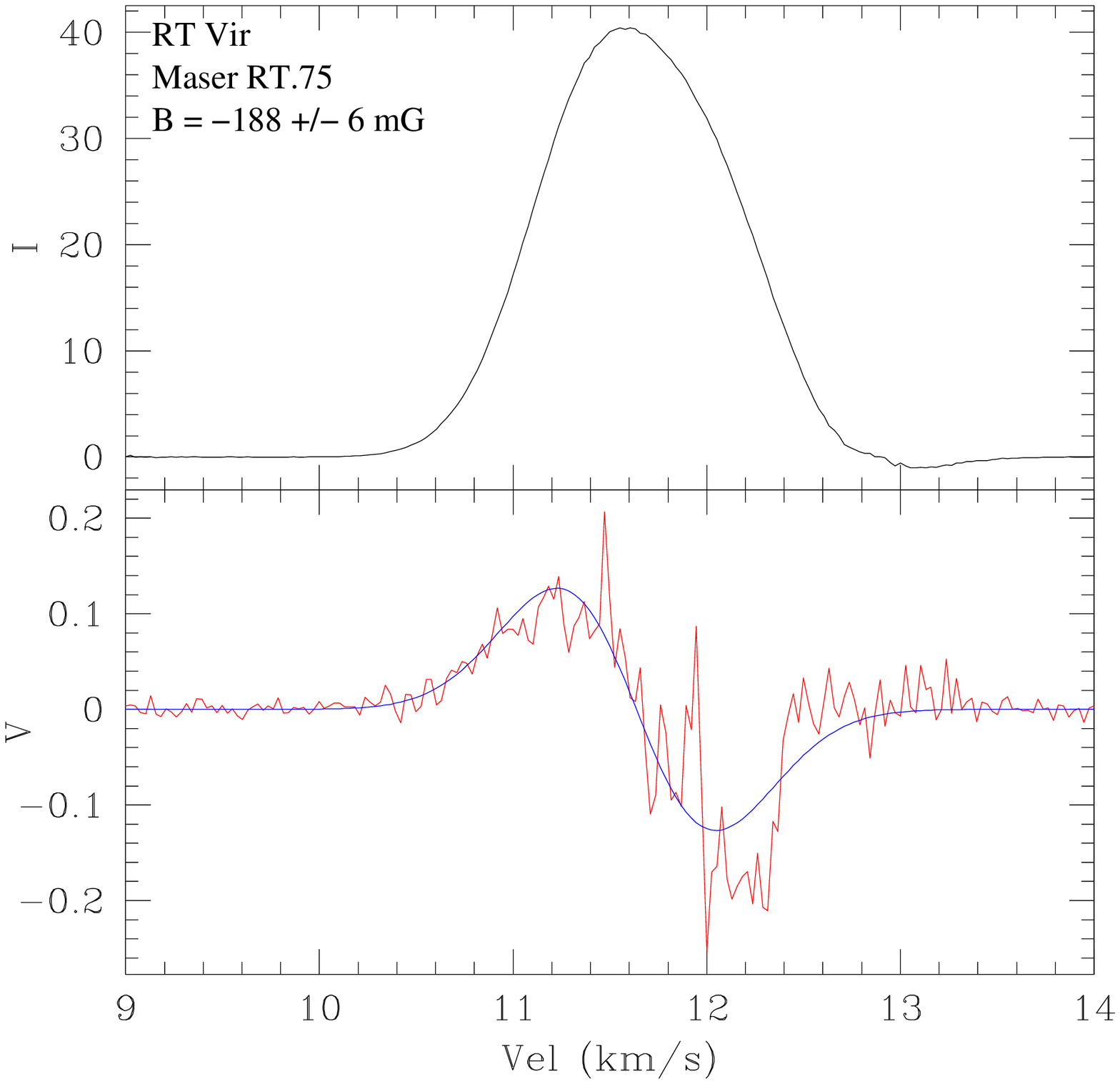} &
  \includegraphics[width = 59 mm]{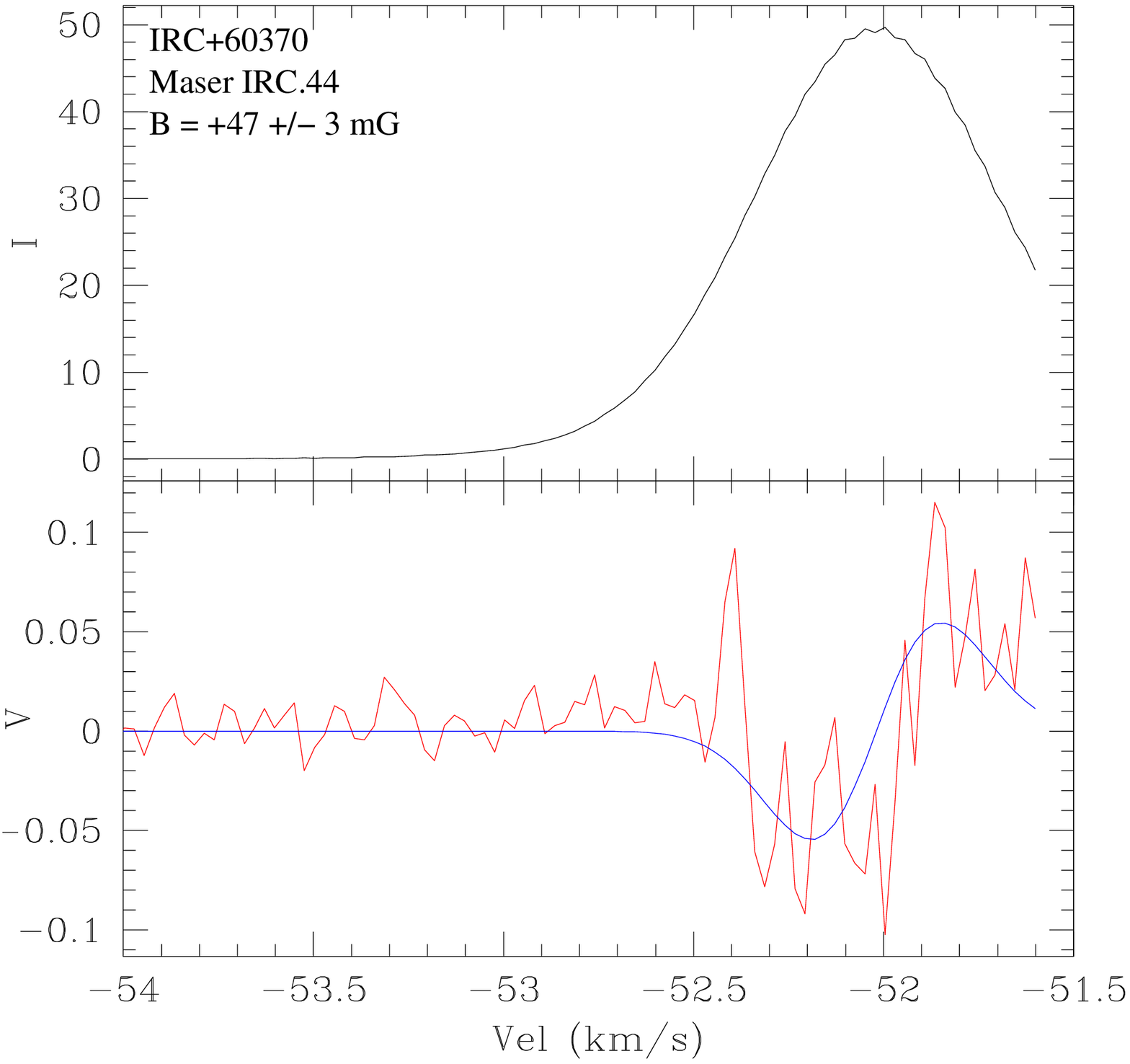} \\
  \includegraphics[width = 59 mm]{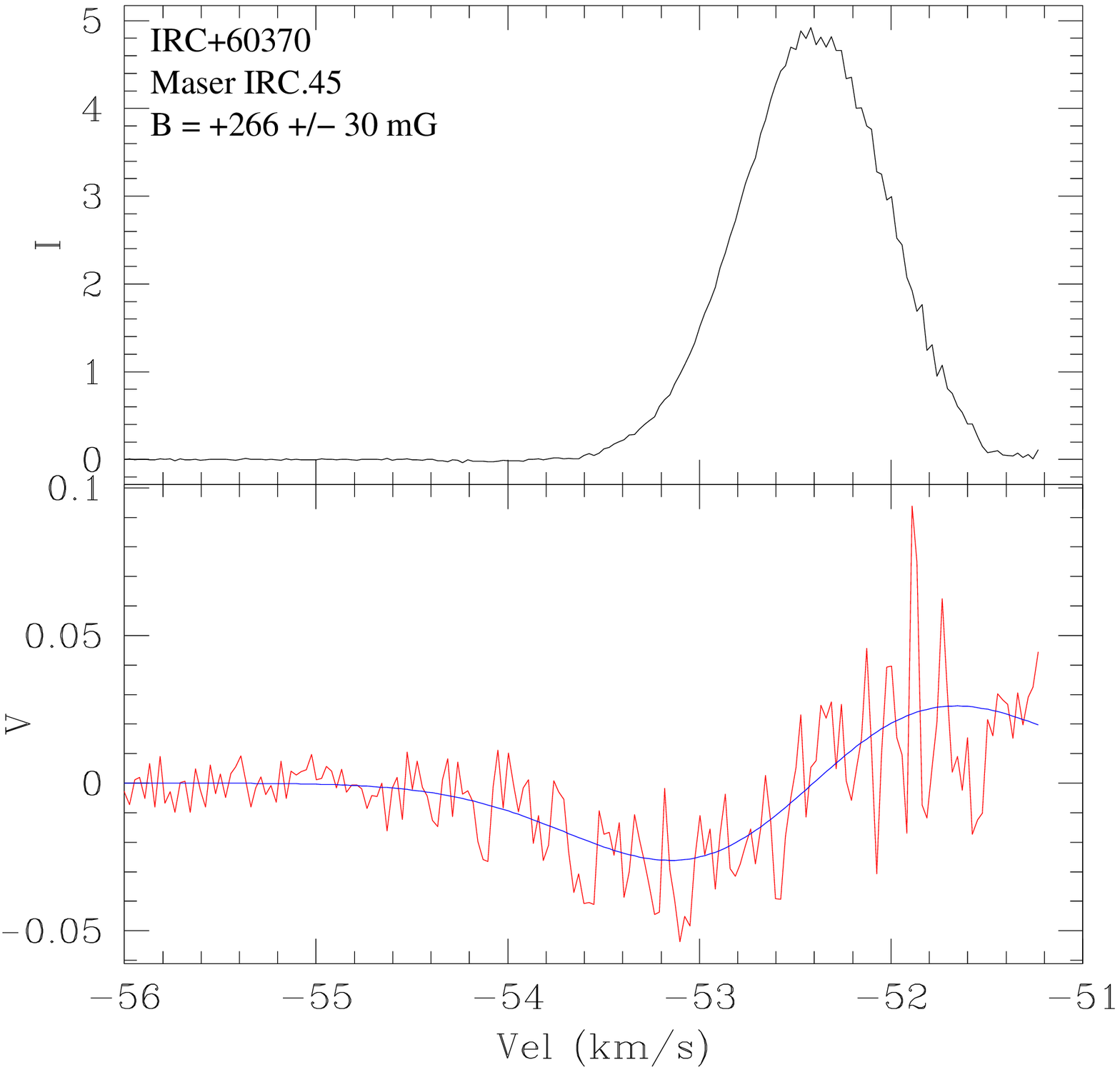} &
  \includegraphics[width = 59 mm]{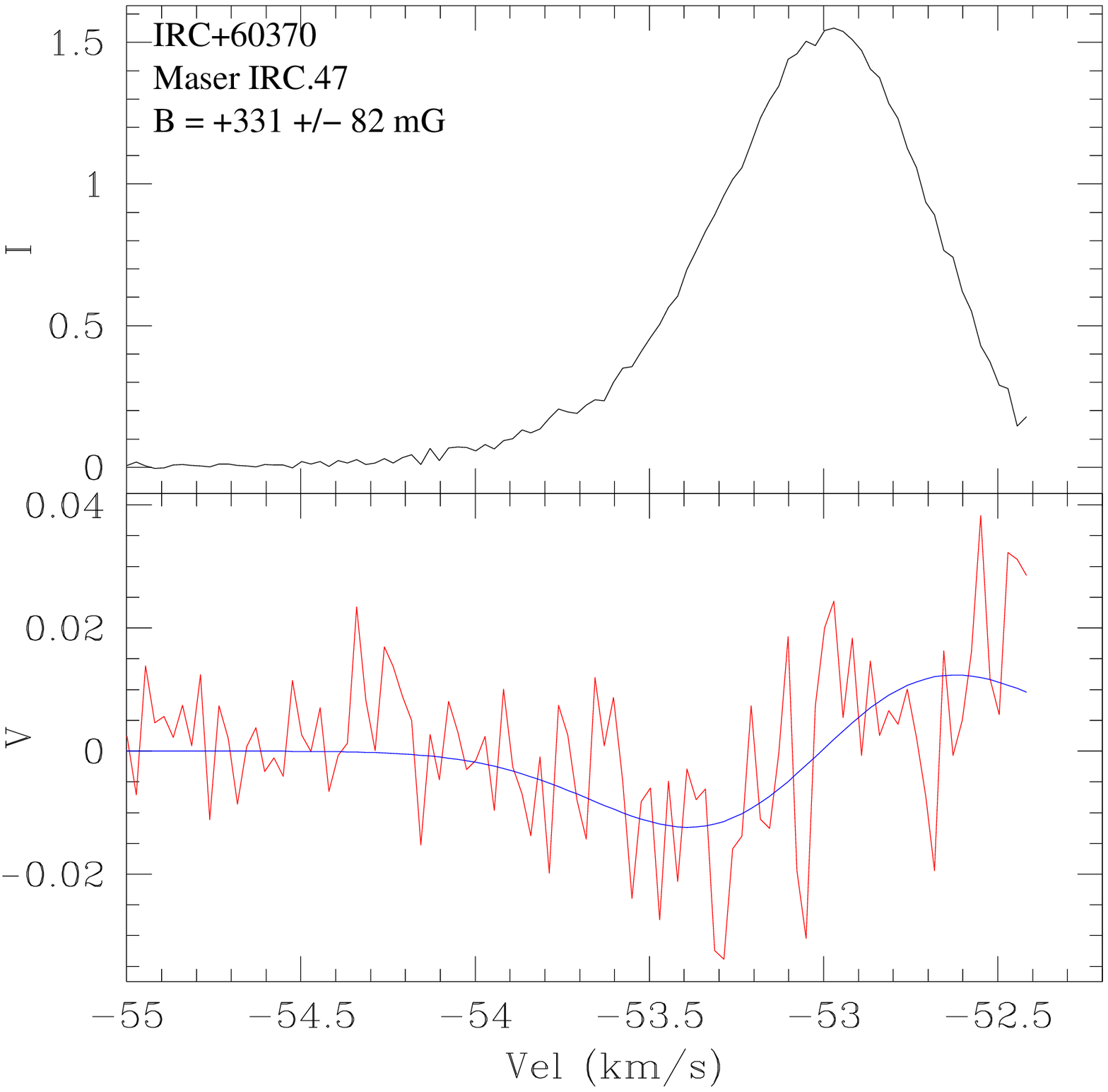} &
  \includegraphics[width = 59 mm]{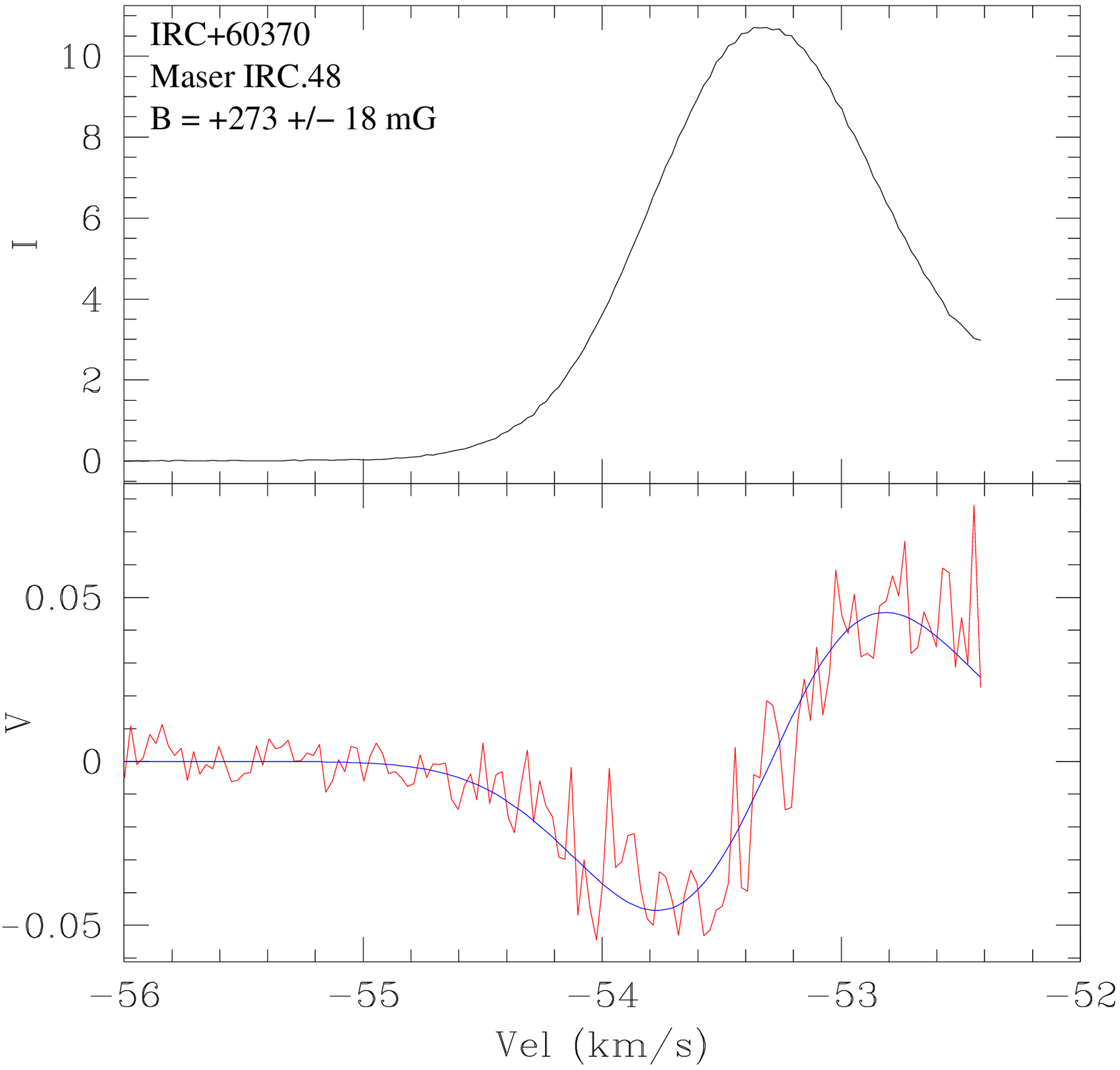} \\
  \includegraphics[width = 59 mm]{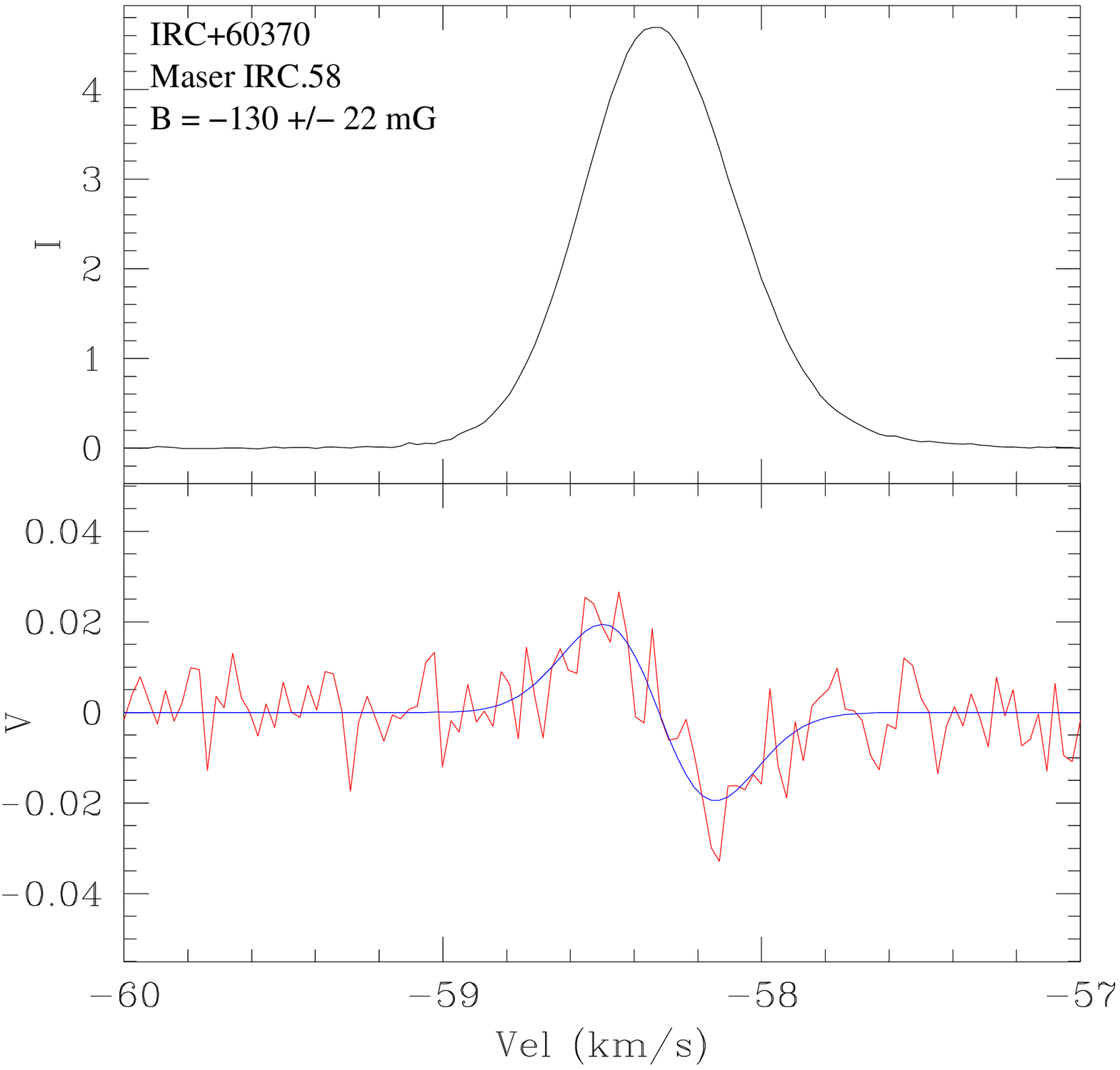} 
\end{tabular}
\caption[]{Plots of Stokes $I$ (top; black line) and $V$ (bottom; red line) 
spectra of all maser features in which we detected circular polarization. The 
blue lines show the best model fit to each $V$ spectrum. The source, the 
maser identification, and the magnetic field strength along the line of sight 
given by the fit are presented in top-left corner of each plot. The 
x-axis shows $V_{LSR}$ in km/s and the y-axis the intensity in Jy/beam.}
  \label{circpol}
\end{figure*}

We observed a total of 642 \h2o maser spots toward IK~Tau. Of these, 525 spots 
survived the multi-channel criteria and comprise 85 maser features around 
this source. In Figs.~\ref{masermaps}.I, we present the spatial distribution 
of these 85 maser features. In Fig.~\ref{masermaps}.II and~\ref{masermaps}.III, 
we zoom in on the two areas indicated in Fig.~\ref{masermaps}.I.

We did not find linear polarization in any maser feature around IK~Tau. 
However, circular polarization was detected in three features around this 
source: IK.20, IK.69, and IK.84 (see Table~\ref{results}). The magnetic 
field strength along the line of sight given by the model fits are: 
$-$147$\pm$15~mG, $-$96$\pm$31~mG, and $+$215$\pm$56~mG, respectively. 
These features are identified in Fig.~\ref{masermaps} labeled according to 
their field strengths.

\subsection{RT~Vir}
\label{rtvir}

We observed 830 \h2o maser spots toward RT~Vir in total. Of these, 671 spots 
comprise 91 maser features around this source. In Fig. \ref{masermaps}.IV, 
we present the spatial distribution of these 91 maser features and in Fig. 
\ref{masermaps}.V we show an enlargement of the area indicated in Fig. 
\ref{masermaps}.IV.

We detected linear polarization in nine features toward RT~Vir: RT.31, RT.34, 
RT.67, RT.68, RT.70, RT.72, RT.73, RT.75, and RT.90 (see Table~\ref{results}). 
Unfortunately, no good polarization calibrator was available, making it 
impossible to determine the absolute direction of the polarization vectors 
(the relative $EVPA$ between components is still correct). 

The distribution of $EVPA$ among the nine features clearly distinguishes 
two groups of masers. Six features, located within projected right ascension 
offset $-$40~$\lesssim\alpha_{off}\lesssim$~0~mas and declination offset 
$-$10~$\lesssim\delta_{off}\lesssim$~30~mas (Fig.~\ref{masermaps}.V) have 
$EVPA$ between $-$38$^\circ$ and $-$59$^\circ$. Another group of features, 
located within 25~$\lesssim\alpha_{off}\lesssim$~35~mas and 
$-$15~$\lesssim\delta_{off}\lesssim$~5~mas (Fig.~\ref{masermaps}.V), also has 
a small $EVPA$ dispersion ($+$38$^\circ$~$\leq$~$EVPA$~$\leq$~$+$64$^\circ$).

Circular polarization was found in three features around RT~Vir: RT.70, RT.75, 
and RT.90. From the fit of the $V$ spectra, we inferred magnetic field 
strengths along the line of sight of $-$143$\pm$12~mG and $-$188$\pm$6~mG in 
RT.70 and RT.75, respectively. We note, however, that the model fit of RT.70 
does not superimpose the whole S-shape structure of its V spectrum. This is 
a consequence of the strong emission that peaks at 11.7~km/s (RT.75). Because 
of this strong emission, a higher noise is present in the spectra around 
11.7~km/s. Therefore, we truncated the RT.70 spectrum at velocity values lower 
than 12.6~km/s to minimize the impact of this noise on the fit. 
However, even with this truncation, a high noise is still present in part 
of the V spectrum and so the results from the model fit of RT.70 should 
be taken with caution. 

\begin{figure}
 \centering
 \begin{tabular}{cc}
   \includegraphics[width = 73 mm]{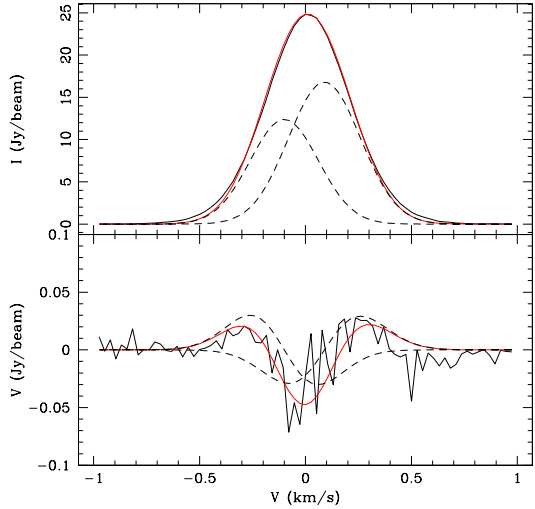}
 \end{tabular}
 \caption[]{Spectra I and V (black curves) of RT.90. The dashed lines show 
the fit of the individual blended features. In red, we show a fit for I and V.}
   \label{rt90}
 \end{figure}

The shape of the $V$ spectrum of RT.90 suggests blended emission. There 
are many free parameters to be taken into account in fitting emission of 
blended features. Consequently, any attempt to obtain a magnetic field 
strength from RT.90 will not generate a solution that is unique or robust. 
However, it is important to emphasize that the shape of its $V$ spectrum 
clearly indicates the presence of a magnetic field. As an example, we created 
a possible fit for this feature. The solution we found for this fit gives 
a magnetic field of $-$84~mG for the slightly more blue-shifted emission 
and $+$63~mG for the slightly more red-shifted feature. The features 
themselves are separated by approximately 0.2~km/s and have widths of 
0.38 and 0.4~km/s. We present this possible fit in Fig.~\ref{rt90}. 
 
In Fig.~\ref{masermaps}.V, RT.70, RT.75, and RT.90 are labeled with the 
magnetic field strength along the line of sight obtained from the model 
fits shown in Figs.~\ref{circpol} and \ref{rt90}.

\subsection{IRC$+$60370}
\label{irc60370}

We observed 658 \h2o maser spots in IRC$+$60370 in total. Of these, 634 spots 
comprise 62 maser features around this source. In Fig.~\ref{masermaps}.VI, we 
present the spatial distribution of these 62 maser features and in 
Fig.~\ref{masermaps}.VII and \ref{masermaps}.VIII we show an enlargement of 
the areas indicated in Fig.~\ref{masermaps}.VI.

We detected linear polarization in nine features toward IRC$+$60370. 
These nine features are concentrated in two small projected regions, 
with areas of $\sim$2~mas$^2$ each. Six of them are located within 
$-$2~$\lesssim\alpha_{off}\lesssim$~0~mas and 
$-$1~$\lesssim\delta_{off}\lesssim$~3~mas (Fig.~\ref{masermaps}.VII; 
Zoom A). The other three features with linear polarization detection 
are located within 27~$\lesssim\alpha_{off}\lesssim$~29~mas and 
6~$\lesssim\delta_{off}\lesssim$~8~mas (Fig. \ref{masermaps}.VIII; Zoom B). 
The EVPA of the masers in Zoom A vary from $-$97$^\circ$ to $-$39$^\circ$, while 
the EVPA of the masers in Zoom B are between $-$74$^\circ$ and $-$132$^\circ$.

Circular polarization was found in five features around IRC$+$60370: IRC.44, 
IRC.45, IRC.47, IRC.48, and IRC.58. From the model fit to the V spectra, we 
measured magnetic field strengths along the line of sight of $+$47$\pm$3~mG, 
$+$266$\pm$30~mG, $+$331$\pm$82~mG, $+$273$\pm$18~mG, and $-$130$\pm$22~mG, 
respectively. We note that, once more, the model fit of several features does 
not superimpose the full expected S-shape of the V spectra. For this source 
three factors contributed to this: (i) the limit of the observed spectral 
range, (ii) features with similar spatial and spectral position, and (iii) 
the increase in the noise near -52 km/s, due to the strong feature IRC.44. 
Therefore, the results given by the fit to IRC.44, IRC.45, IRC.47, and 
IRC.48 should also be taken with caution. In Fig.~\ref{masermaps}.VII, 
the five features with circular polarization detection are labeled with 
the magnetic field strength along the line of sight that we obtained from 
the model fits shown in Fig.~\ref{circpol}.

\section{Discussion} 
\label{discussion}

\subsection{Non-detection toward AP~Lyn} 
 \label{aplyndisc}

Several \h2o masers toward AP~Lyn have been detected before 
\cite[e.g.,][]{imai97,migenes99,colomer00,sudou02,shintani08,kim10}. 
 \citet{shintani08} monitored this and other sources from 2003 to 
2006, using the Iriki 20~m telescope of the VLBI Exploration of Radio 
Astrometry (VERA). They reported a high flux variation, and fit 
a maser light curve for Ap~Lyn. The peak flux of the \h2o masers 
reported in the other works vary from $\sim$6~Jy \cite[observed 
with the VLBA in 1996;][]{migenes99} up to $\sim$120~Jy \cite[observed 
with the Kashima-Nobeyama Interferometer in 1992;][]{sudou02}. 
Conservatively, we suggest an upper limit of 1~Jy for the flux 
density of our non-detection (in the raw data).

\citet{richards12} give a detailed discussion of the possible 
causes of \h2o maser variability. They point out that a correlation 
of the infrared light curve and maser variability can exist. Also, 
they disagree with previous papers \cite[e.g.,][]{shintani08} stating 
that no systematic relationship between maser brighness and the optical 
phase was found at the times of their observations.

\subsection{Spatial distribution of the masers} 
 \label{spatialdist}

The spatial distribution of the features around IK~Tau shows a clear 
correlation between velocity and position. While masers with higher 
velocities (red circles) are concentrated in the west and southwest, 
the features with lower velocity (blue circles) appear, mostly, in the 
east and northeast (Fig.~\ref{masermaps}.I). This behavior is 
also reported by \citet{bains03}. They suggest that the shell of IK~Tau has 
an equatorial density enhancement. The brightest masers would lie in an oblate 
spheroid and the plane of the equator would have an inclination angle $i'$ 
with the line of sight (45$^{\circ} \lesssim i' \lesssim$ 90$^{\circ}$). The 
eastern end of the polar axis would then be approaching us, explaining 
the east-west velocity segregation. This model also explains why the IK~Tau 
observations made more than 10 years apart show a persistent east-west 
offset between moderately red- and blue-shifted emission, although 
individual masers do not survive for more than $\sim$1.5 years. Our IK~Tau 
data were observed in 2009, almost 15 years later than the observations 
reported by \citet{bains03}, and 24 years after the observations reported 
by \citet{yates94}.

\citet{bains03} also observed a similar east-west velocity offset in RT~Vir. 
Our data do not show a clear correlation between velocity and 
position for this source (Fig.~\ref{masermaps}.I), but a moderate enhanced 
concentration of red-shifted features in the east is present, while the bluer 
features are concentrated in the center of the plot. This is different 
from the east-west relation seen in Fig.~6 of \citet{bains03}. In their 
figure, the red-shifted masers are located on the western side, and the 
blue-shifted features are concentrated on the eastern side. 

An individual \h2o maser has its life time estimated to be less than 
1--2~years. Multi-epoch imaging of 22~GHz \h2o masers often shows major 
changes in the maser distribution over the years \cite[e.g.,][]{richards12}. 
IK~Tau is, therefore, an exception to this behavior.

\subsection{Stellar Position}
 \label{starpos}

Some of the analysis discussed in this paper requires information concerning 
the stellar position in relation to the observed masers (Sects. 
\ref{distance} and \ref{magfield}). However, the absolute stellar 
position is not known for our observations. So to infer the stellar 
position, we used the shell-fitting method \citep{yates93,bains03}. 
This method assumes a distribution of masers on a tridimensional sphere, 
with the star located in its center. All masers in a velocity range 
determined by
\begin{eqnarray}
V_{star} \pm i(\Delta V_{LSR}/n) 
\label{eq2}
\end{eqnarray}

\noindent are identified, where $V_{star}$ is the velocity of the star, 
$\Delta V_{LSR}$ the total maser velocity range, and $n$ is a number taken 
here to be equal to 8. We choose that value to restrict the selection of 
the masers to lie within a small velocity range. The constant $i$ sets which 
ring(s) along the line of sight is considered. If $i$ is equal to 1, then 
a ring at the same line of sight velocity as the star is taken. If $i$ is 
bigger than 1, then one ring in front and one behind of the star are 
considered. Once the masers are selected, the central position of the 
features is assumed to be the stellar position. We emphasize that the 
more asymmetric the maser distribution, the larger the uncertainty of this 
method.

For each object, we varied the value of $i$, obtaining different locations 
for the stellar position. An additional position was calculated by 
taking the center point of all the observed masers. We assumed the stellar 
position to be the mean location of the different positions we obtained by 
using different values of $i$, and by using the center point of all the 
observed masers. In Table~\ref{star} we show the stellar position we 
calculated for each value of $i$ and the mean result.

\begin{table}
  \begin{center}
    \caption{Stellar position}
    \begin{tabular}{@{}clll@{}}
      \hline
             & IK~Tau    & RT~Vir                & IRC$+$60370\\
             & $\alpha$, $\delta$ (mas) & $\alpha$, $\delta$ (mas) & $\alpha$, $\delta$ (mas) \\
      \hline
$i$=1        & --                & $\alpha$ = +04.51 & $\alpha$ = +07.44\\
             & --                & $\delta$ = --13.83& $\delta$ = +03.57\\
$i$=2        & $\alpha$ = +07.43 & --                & $\alpha$ = +14.49\\
             & $\delta$ = +03.09 & --                & $\delta$ = +03.61 \\
$i$=3        & $\alpha$ = +28.58 & --                & $\alpha$ = +11.17\\
             & $\delta$ = --05.38& --                & $\delta$ = +01.99 \\
All Features & $\alpha$ = +10.41 & $\alpha$ = +10.69 & $\alpha$ = +10.92\\
             & $\delta$ = +02.36 & $\delta$ = --15.32& $\delta$ = +02.68\\
             \hline
Mean Position& $\alpha$ = +15.47 & $\alpha$ = +07.60 & $\alpha$ = +11.01 \\
             & $\delta$ = +00.02 & $\delta$ = --14.58& $\delta$ = +02.96 \\
             \hline
    \end{tabular}
    \label{star}
    \end{center}
{Position of the star, relative to the reference maser, for different 
values of $i$. The positions we obtained as the centroid of all the observed 
maser features are also shown. Finally, the mean result is reported 
at the bottom of the table. Columns 2 to 4 show the stellar position 
of IK~Tau, RT~Vir, and IRC$+$60370.}
\end{table}

\subsection{Distance of the masers to the star} 
 \label{distance}

\begin{figure}
  \centering
  \begin{tabular}{cc}
    \includegraphics[width = 78 mm]{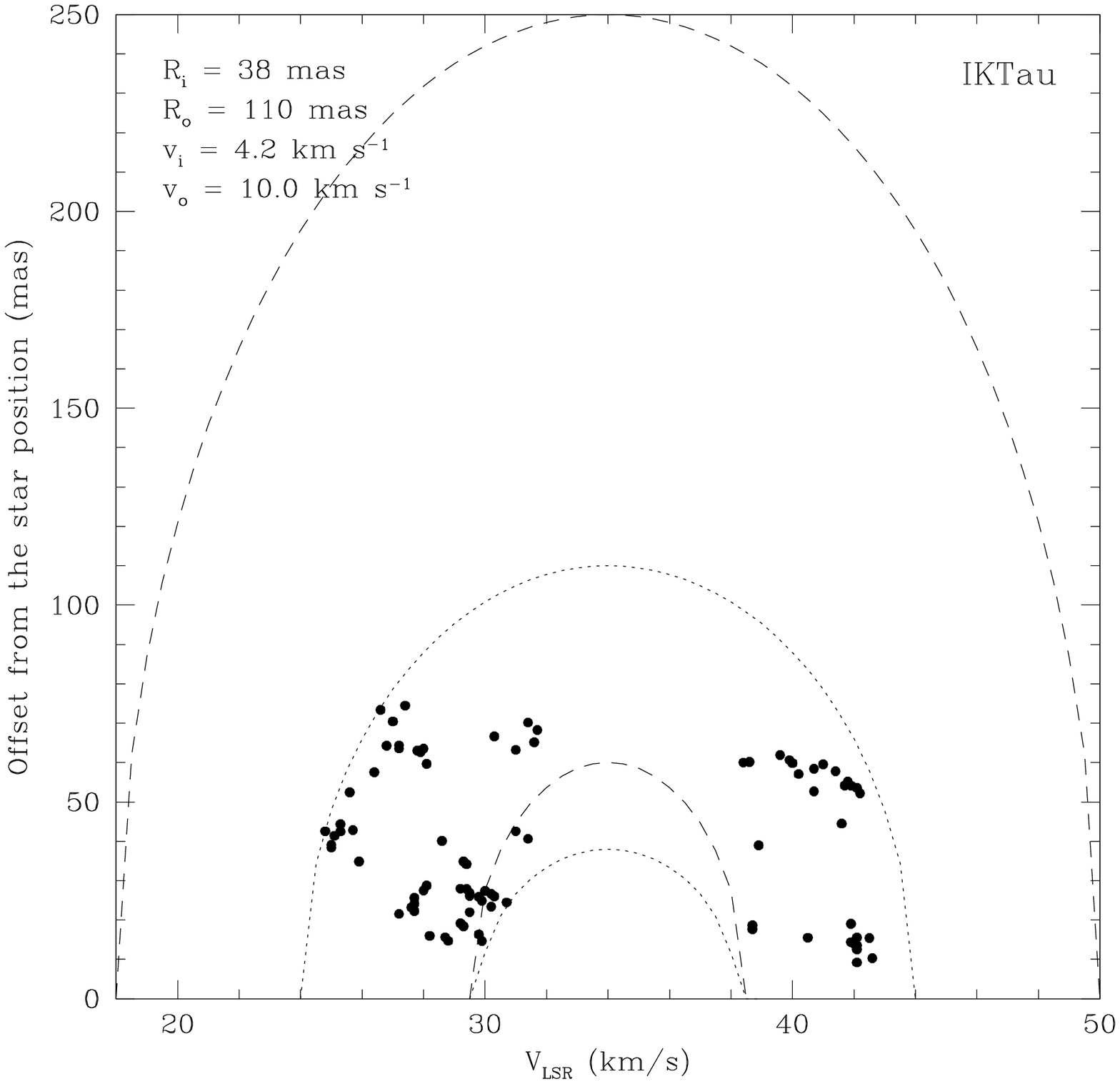} \\
    \includegraphics[width = 78 mm]{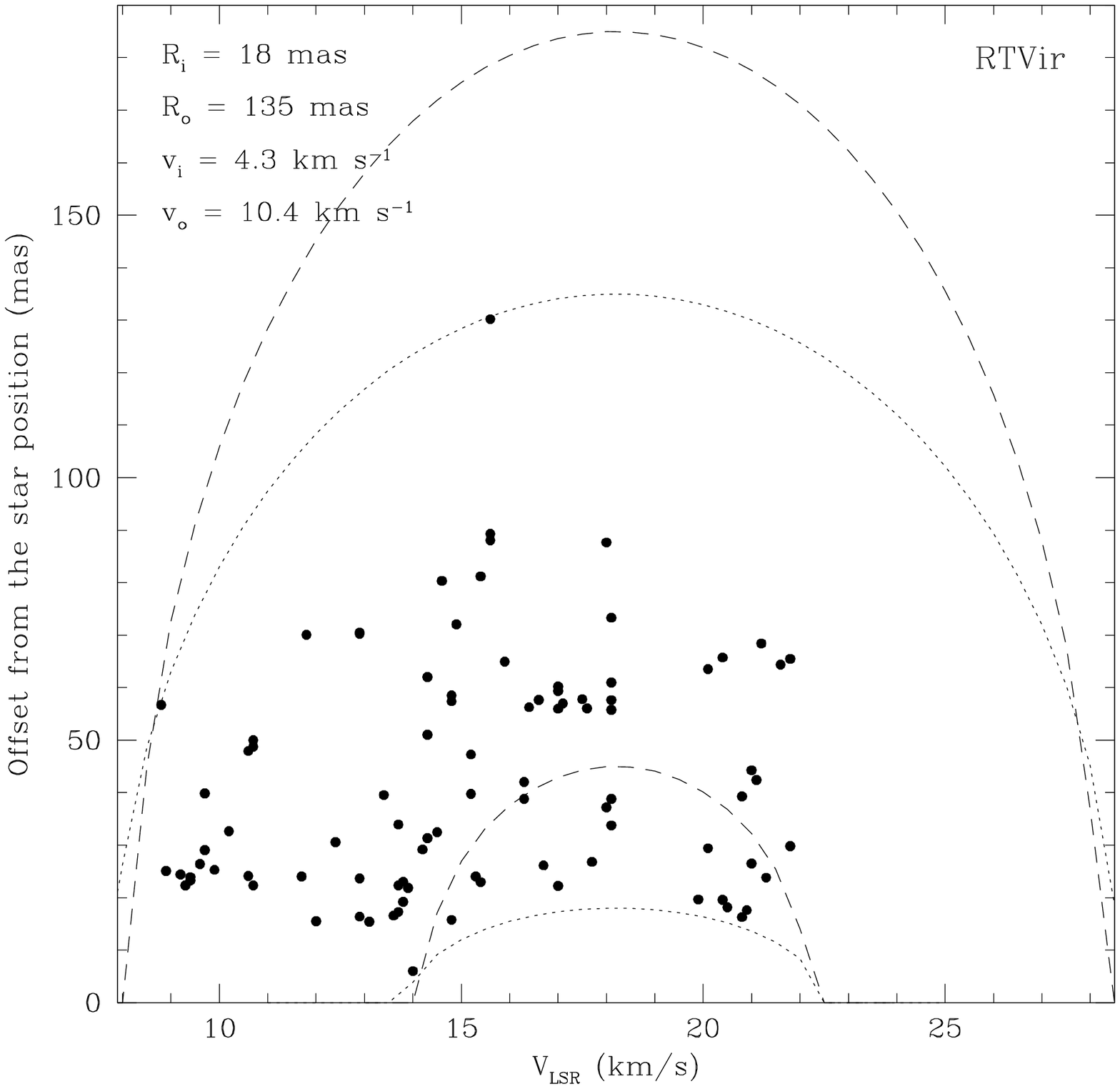} \\
    \includegraphics[width = 78 mm]{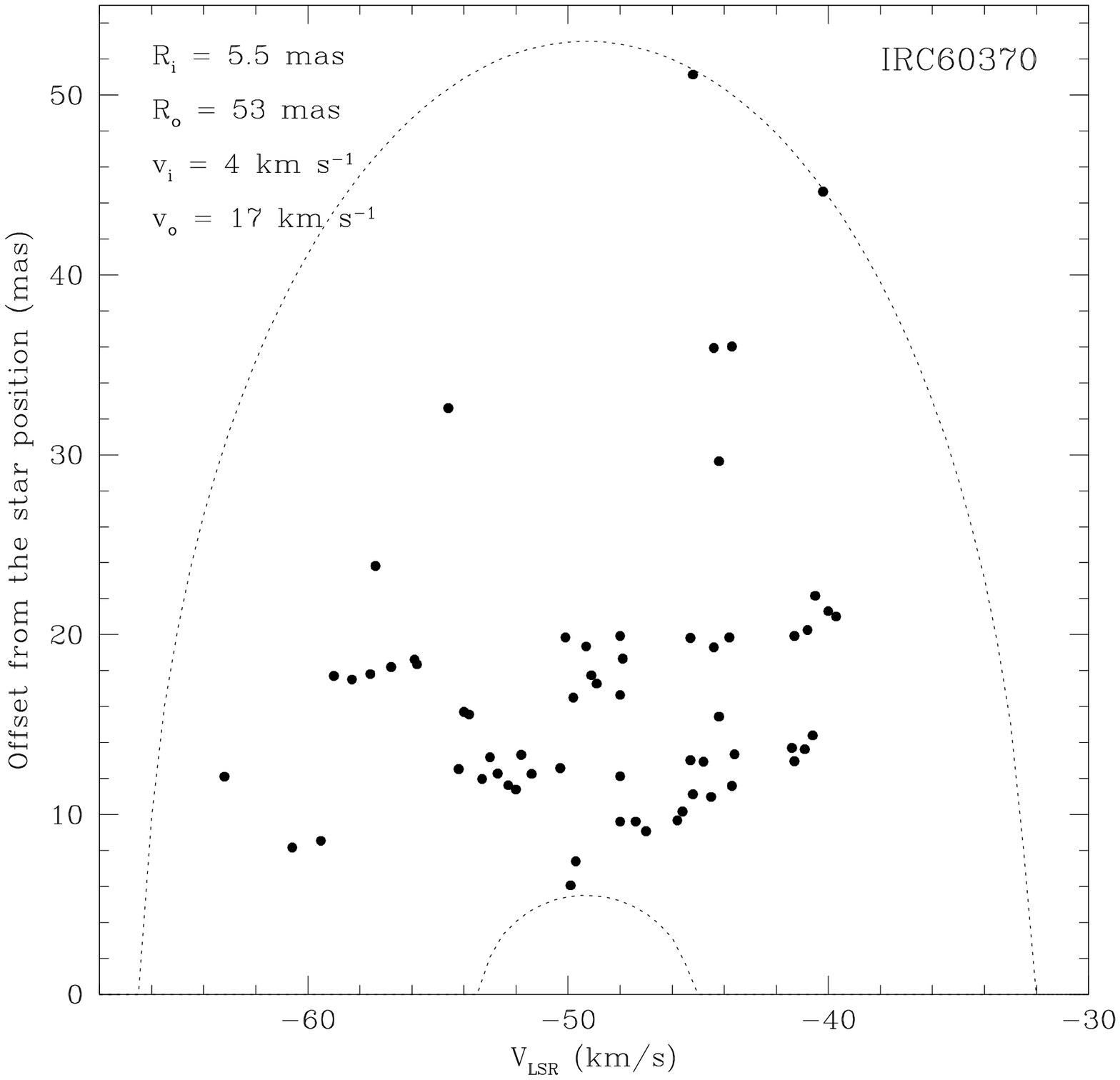}
  \end{tabular}
  \caption[]{Velocity of the features versus their projected offsets from the 
    star. The dotted lines represent our fits; the dashed lines were 
    reproduced from \cite{bains03}. From top to bottom, we show the plots for 
    IK~Tau, RT~Vir, and IRC$+$60370.}
  \label{distvel}
\end{figure}

In Fig. \ref{distvel} we show, for each source, a plot of the velocity of the 
features versus their projected angular offsets from the star ($\theta_{off}$; 
see Sect.~\ref{starpos} for the determination of the stellar position). 
For each source, two parabolas are fitted to the velocity-offset positions. 
These fits are shown by the dotted lines in the figures. In the fitting 
process, made by eye, the area between the parabolas which contains 
all masers is minimized. The parabolas obey the relation 
\begin{eqnarray}
\theta_{off} &=& \frac{R}{V_{shell}} \times (V_{shell}^2 - (V_{LSR}-V_{star})^2)^{1/2}, 
\label{eq3}
\end{eqnarray}
\noindent where $R$ is the distance to the star, $V_{shell}$ the expanding 
velocity of the masers, and $V_{star}$ the velocity of the star. 

Assuming that the masers are located in a spherical shell around the star, 
it is possible to determine the internal ($R_i$) and external ($R_o$) radius 
of this shell from the internal and external parabola fits, and their 
corresponding expansion velocities ($v_i$ and $v_o$). The values we adopted 
for $V_{star}$, the distance to the source, their respective references, 
and the fit parameters ($v_i$, $v_o$, $R_i$, $R_o$) are shown in 
Table~\ref{in/output}. 

\citet{bains03} and \citet{richards11} also investigated the kinematics of 
IK~Tau and RT~Vir and found similar results for $v_i$, $v_o$, $R_i$, and $R_o$. 
To illustrate the comparison with our results, we reproduce the fits 
from \cite{bains03} for IK~Tau and RT~Vir in our Fig.~\ref{distvel}. Those 
authors present two alternative solutions for the internal fit to IK~Tau. 
We choose to show only the one with the larger radius here. Their fits 
are shown in Fig.~\ref{distvel} by the dashed lines. We note that there 
is a big disagreement between the external fits from \cite{bains03} and 
ours. This is probably because our observations with the VLBA resolve 
out more diffuse emissions, due to its longer baselines. Additionally, 
our result implies that the \h2o maser regions around IK~Tau and RT~Vir 
reach closer to the star than was determined by \cite{bains03}. 
Quantitatively, we found $R_i$ equal to 38 and 18~mas for IK~Tau and 
RT~Vir, respectively. The fits that we reproduced from \cite{bains03} 
correspond to $R_i$ equal to 60 and 45~mas for IK~Tau and RT~Vir, 
respectively. We emphasize, however, that their alternative solution 
for the internal fit of IK~Tau shows an inner radius of the \h2o maser 
region closer to the star than ours ($R_i$ equal to 25~mas). For IK~Tau, 
\cite{richards11} found $R_i$ between 60 and 75~mas for different epochs, but 
they also detected a faint group of masers with $R_i$ smaller than 64~mas 
(at 23~mas). For RT~Vir, \cite{richards11} found $R_i$ between 34 and 
45~mas for different epochs. Hence, considering the stellar radius of 
IK~Tau and RT~Vir to be, respectively, 0.8~AU and 2.8~AU \citep{monnier04,
ragland06,richards12}, it seems that although the majority of the 22~GHz 
\h2o masers occur outside a distance of $\sim$5--7 stellar radii, occasional 
clumps can be found as close as $\sim$3 stellar radii.

\begin{table*}
  \begin{center}
    \caption{Distance of the masers to the star: input and output parameters}
    \begin{tabular}{@{}lcccccccc@{}}
      \hline
Source&$V_{star}$ (ref)& $D$ (ref)&$v_i$&$v_o$& $R_i$ & $R_o$ & $R_i$ & $R_o$\\
&(km/s)& (pc) & (km/s) & (km/s) & (mas) & (mas) & (AU) & (AU) \\
      \hline
IK~Tau     & $+$34.0 (K87) & 265 (H97) & 4.2 & 10.0 & 38 & 110 & 10.1 & 29.2 \\
RT~Vir     & $+$18.2 (N86) & 133 (H97) & 4.3 & 10.4 & 18 & 135 &  2.4 & 18.0 \\
IRC$+$60370& $-$49.3 (I08) & 1000 (I08)& 4.0 & 17.0 & 5.5&  53 &  5.5 & 53.0 \\
             \hline
    \end{tabular}
    \label{in/output}
  \end{center}
{From Cols. 1 to 9: the source name (Source), the velocity of the source 
and its reference ($V_{star}$ (ref)), the distance 
to the source and its reference ($D$ (ref)), the inner ($v_i$) and outer 
($v_o$) expansion velocities of the \h2o envelope, and the inner ($R_i$) and 
outer ($R_o$) distances of the \h2o maser region to the star, both in mas 
and AU.  References: K87: \cite{kirrane87}; N86: \cite{nyman86}; I08: 
\cite{imai08}; H97: \cite{hipparcos97}.}
\end{table*}

\subsection{Magnetic field detection} 
 \label{magfield}

\subsubsection{Linear polarization and field geometry} 
 \label{linpol}

We measured fractional linear polarization from 0.1$\%$ to 1.4$\%$ around 
RT~Vir, and between 0.2$\%$ and 1.6$\%$ around IRC$+$60370. The non-detection 
of linear polarization in the features around IK~Tau imply that, if 
present, it is lower than $\sim$0.5$\%$ on the brightest masers. 
These results agree with the upper limits for fractional linear 
polarization derived from the non-detections of \cite{vlemmings02}. 

According to maser theory, the magnetic field lines can be either 
parallel or perpendicular to the EVPA. It is parallel when the angle 
$\theta$ between the field and the direction of propagation of the maser 
is less than the Van Vleck angle ($\sim$55$^{\circ}$), and perpendicular 
when $\theta$ is greater than the Van Vleck angle \citep{goldreich73}. 
The linear polarization is affected by $\theta$ and the degree of 
saturation but, based on our measured values, we cannot ensure in 
which regime - parallel or perpendicular - the emission originates.

As shown in Sects. \ref{rtvir} and \ref{irc60370}, linear polarization 
has been detected in masers toward RT~Vir and IRC$+$60370. In each of these 
sources, the polarized features are separated into two groups. In RT~Vir, 
both of these groups show a small EVPA dispersion ($\leq$ 26$^{\circ}$ for all 
masers within a given group). If, in this source, we are dealing with 
a magnetic field perpendicular to the EVPA, either a poloidal or a dipole 
field seems to be the best qualitative fit of the field geometry to our 
results. On the other hand, if the field is parallel to the EVPA, the 
polarization vectors could trace tangent points of a toroidal field. 
In IRC$+$60370, the EVPA of the features have a higher dispersion, but 
the vectors still seem to trace a dominant direction, pointing towards the 
position of the star, especially in the features located within the Zoom~A 
region. If, in this source, we are dealing with a magnetic field perpendicular 
to the EVPA, either a poloidal or a dipole field could be argued as 
probable fits to our results. On the other hand, if the field is parallel 
to the EVPA, a toroidal field may provide a better qualitative fit. 
Furthermore, we detected circular polarization in four features located 
within the Zoom~A region and, from the model fit of their V~spectra, all 
of them show a magnetic field strength with a positive sign. Inside 
the Zoom~B region, however, the single feature in which we detected 
circular polarization shows a magnetic field strength with a negative 
sign. These results lead to the conclusion that the component of the 
magnetic field along the line of sight points in opposite directions on either 
side of the star. That evidence suggests, again, a toroidal field around 
IRC$+$60370.

\subsubsection{Magnetic field dependence} 
 \label{Bdepend}

\begin{figure}
 \centering
 \begin{tabular}{cc}
   \includegraphics[width = 90 mm]{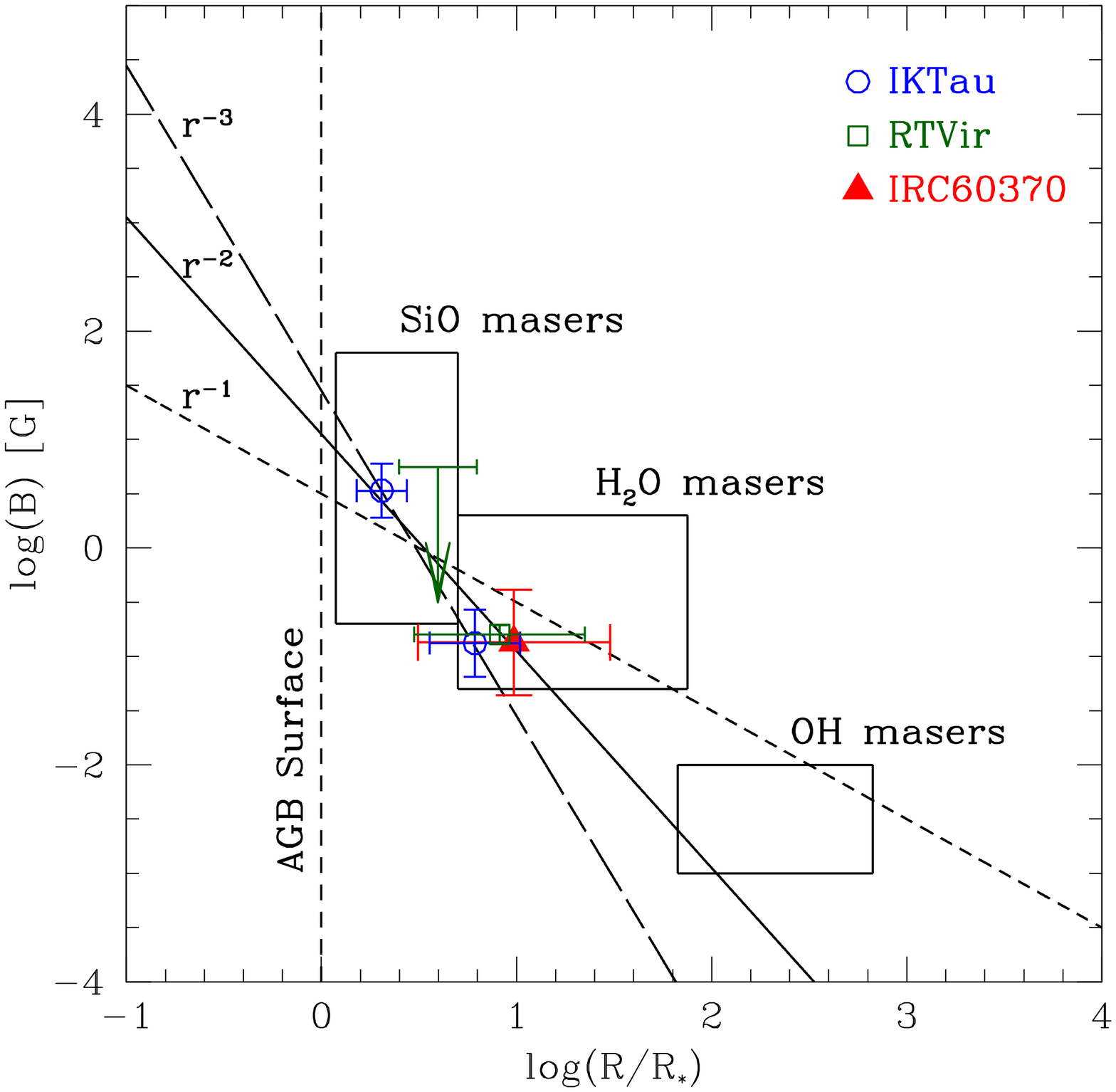} \\
 \end{tabular}
 \caption[]{Magnetic field strength along the line of sight versus the 
   radial distance of the masers to the star. The black boxes show typical 
   regions of the plot where results from the literature for SiO, \h2o, and 
   OH maser occur, and they are normalized for R$_*$=1 
   \citep{vlemmings02,vlemmings05,herpin06,rudnitski10}. 
   Our measurements are shown by the hollow blue circles (IK~Tau), hollow 
   green squares (RT~Vir), and filled red triangles (IRC$+$60370). 
   The short-dashed, solid, and long-dashed inclined lines show a 
   dependence $\propto$~$R^{-1}$, $\propto$~$R^{-2}$, and $\propto$~$R^{-3}$ 
   for the magnetic field, respectively. The position of the AGB surface 
   of a star with radius of 1~AU is also shown.}
   \label{distB}
 \end{figure}

In Fig.~\ref{distB}, we show a plot of the magnetic field strength along the 
line of sight for the stars in our sample, estimated from different maser 
species, against the radial distance of these masers to the star. We use this 
plot to investigate the field dependence on $R$: $B~\propto$~$R^{-\alpha}$, 
where $\alpha$ depends on the structure of the magnetic field in the 
circumstellar envelope. When $\alpha$ equal to 1, it refers to a toroidal 
magnetic field, $\alpha$ equal to 2 corresponds to a poloidal field, and 
$\alpha$ equal to 3 indicates a dipole geometry. In the plot we show one 
single box where the results of OH masers occur. However, we emphasize that the 
1665/7~MHz OH maser emission originates in inner regions when compared to the 
1612~MHz OH maser transition. Therefore, it is expected that magnetic field 
strength measurements based on the first line to be stronger than the second 
\citep{wolak12}.

In this plot, we included polarization results of the SiO maser 
region from the literature. We took the magnetic field strength in the SiO 
maser region from \cite{herpin06} for RT~Vir (upper limit) and IK~Tau. 
For IK~Tau, the distance of the SiO region to the star was adapted from 
\cite{boboltz05}, adopting a distance to the source of 265~pc. For 
RT~Vir, we used a typical value for the radial distance of the SiO maser region 
(between 2 and 5~AU from the star). Unfortunately, we did not find any 
reports of the magnetic field strength in the OH maser region that would 
allow us to make a more complete plot. For all cases, the major uncertainty 
in the plot concerns $R$.

The data from RT~Vir and IRC$+$60370 do not allow a definitive conclusion 
regarding the functional form of radial dependence. For IK~Tau, however, 
even though a $B~\propto$~$R^{-1}$ dependence is not totally ruled out, 
$B~\propto$~$R^{-2}$ and $B~\propto$~$R^{-3}$ provide qualitatively better 
fits.

\subsubsection{Magnetic Field on the Star Surface} 
 \label{Bstar}

By assuming a magnetic field dependence ($B~\propto$~$R^{-1}$, 
$B~\propto$~$R^{-2}$, or $B~\propto$~$R^{-3}$; see Sect.~\ref{Bdepend}), 
we can extrapolate the projected field strength to the surface of 
the star ($B_{star}$). If $B~\propto$~$R^{-\alpha}$, then 

\begin{eqnarray}
B_{star}~=~B_{H_2O}~\times~(R/R_*)^{\alpha},
\label{eq4}
\end{eqnarray}

\noindent where $B_{H_2O}$ is the field strength along the line of sight 
in the water maser region, and $R_*$ is the stellar radius 
\citep{reid79,reid90}. However, we emphasize that the magnetic 
field could deviate from any power law if the various masers come 
from conditions with different densities, fractional ionization, etc. 
These differences in the physical conditions of the medium could lead 
to differences in how much the field is frozen in, dissipated, 
enhanced by shocks, etc. Therefore, a homogeneous medium is assumed in 
this extrapolation. 

In the analysis with Eq.~\ref{eq4} we investigate each source 
individually, by varying the power law, with $\alpha$ from 1 to 3. For 
IK~Tau and RT~Vir, we adopted $R_*$ as updated by \cite{richards12}. For 
IRC$+$60370, however, there is no accurate measurement for the stellar 
radius. Therefore, for this source we adopted $R_*$=1.8~AU as an upper 
limit based on 18~$\mu$m imaging \citep{meixner99}. 

In order to define the value of $R$ to be given as input in Eq.~\ref{eq4}, 
for each source we created an alternative plot of velocity versus position 
offset (analogous to the procedure described in Sect.~\ref{distance} and 
Fig.~\ref{distvel}). In these alternative plots we considered only the 
features in which we detected circular polarization, getting alternative 
values for $R_i$ and $R_o$ ($R_i'$ and $R_o'$). We adopted $R_i'$ and 
$R_o'$ as minimum and maximum values of $R$ to be given in Eq.~\ref{eq4}. 
We emphasize that these alternative plots were created with very few data 
points, and thus provide only approximate results for $R_i'$ and $R_o'$. We 
combined $R$~=~$R_i'$ with the lowest value of $B_{H_2O}$ that we observed 
(taking the error bar into account -- $B_{H_2O_{min}}$) to derive the lower 
limit of field strength on the surface of the star ($B_{star_{min}}$). For 
the upper limit ($B_{star_{max}}$), we combined $R$~=~$R_o'$ with the highest 
value of $B_{H_2O}$ that we observed (taking the error bar into account -- 
$B_{H_2O_{max}}$).

In Table~\ref{tabBstar}, we show the values given as input in 
Eq.~\ref{eq4} ($R_*$, $B_{H_2O_{min}}$, $B_{H_2O_{max}}$, $R_i'$, and $R_o'$),  
and the results of $B_{star_{min}}$ and $B_{star_{max}}$ for each source.

\begin{table*}
  \begin{center}
    \caption{Projected magnetic field strength on the star surface}
    \begin{tabular}{@{}lcccccccccccc@{}}
      \hline
Source&$R_*$&$B_{H_2O_{min}}$&$B_{H_2O_{max}}$& $R_i'$& $R_o'$&$B_{star_{min}}$&$B_{star_{max}}$&$B_{star_{min}}$&$B_{star_{max}}$& $B_{star_{min}}$& $B_{star_{max}}$\\
& (AU)     &(mG)      &(mG)         & (AU)    & (AU)  &(G)       &(G)   &(G)  &(G)  & (G)& (G)\\
&&&&&& ($\alpha$=1)& ($\alpha$=1)& ($\alpha$=2)& ($\alpha$=2)& ($\alpha$=3)& ($\alpha$=3)\\
      \hline
IK~Tau     &2.8& 65&271&22.5&27.8&0.5&2.7&4.2& 26.8    &33.8&$\sim$265\\
RT~Vir     &0.8&131&194&6.0 &14.6&1.0&3.5&7.3& 64.9    &54.9&$\sim$1185\\
IRC$+$60370&1.8& 44&413&12.5&29.5&0.3&6.9&2.2&$\sim$115&15.8&$\sim$1945\\
             \hline
    \end{tabular}
    \label{tabBstar}
  \end{center}
{From Cols. 1 to 12: the source name (Source), the stellar radius 
($R_*$), the lower ($B_{H_2O_{min}}$) and upper ($B_{H_2O_{max}}$) magnetic field 
strengths along the line of sight observed in the \h2o region, the input 
values of $R$ on Eq.~\ref{eq4} ($R_i'$ and $R_o'$), and the lower 
($B_{star_{min}}$) and upper ($B_{star_{max}}$) limits of the projected magnetic 
field strength on the stellar surface assuming $\alpha=1$, $\alpha=2$, and 
$\alpha=3$.} 
\end{table*}

\subsubsection{Magnetic field energy} 
 \label{Benergy}

One question that needs to be answered to improve our understanding on 
low and intermediate mass stellar evolution is: if present, does the magnetic 
field around evolved stars have enough energy to drive the shaping of these 
objects? 

Several magnetic field detections around AGB and post-AGB stars have been 
reported in recent years \cite[e.g.,][]{amiri11,perez11,lf12,vlemmings12}. 
To check if the magnetic energy density ($U=B^2/2\mu_0$) is important, we 
compare it with typical values of the kinetic and thermal energy density 
around evolved stars (Table~\ref{tabBenergy}). 
For the calculation of these values we assume $V_{exp}$$\sim$5~km/s, 
$n_{H_2}$$\sim$10$^{14}$~cm$^{-3}$, and $T$$\sim$2500~K at the stellar 
photosphere, and $V_{exp}$$\sim$8~km/s, $n_{H_2}$$\sim$10$^{8}$~cm$^{-3}$, and 
$T$$\sim$500~K in the \h2o maser region. In Table~\ref{tabBenergy}, we show 
the limits of the magnetic energy density in the \h2o maser region that we 
observed and the magnetic energy density extrapolated to the surface of the 
star. The limits are based on the field strengths along the line of sight 
reported in Table~\ref{tabBstar}. Our results show that the magnetic 
energy density is dominant in the \h2o maser region. Therefore, the magnetic 
fields probably play an important role in shaping the different morphologies 
of evolved stars that are progenitors of PNe. 

The dominant energy on the surface of the star is still inconclusive 
since different conclusions can be drawn if a dependence of either 
$B~\propto$~$R^{-1}$, $B~\propto$~$R^{-2}$, or $B~\propto$~$R^{-3}$ 
is assumed.

\begin{table}
  \begin{center}
    \caption{Magnetic energy density}
    \begin{tabular}{@{}lcccc@{}}
      \hline
      Source&$U_{H_2O}$&$U_{Star}$ ($\alpha$=1)&$U_{Star}$ ($\alpha$=2)&$U_{Star}$ ($\alpha$=3)\\
      &(J/m$^3$)&(J/m$^3$)&(J/m$^3$)&(J/m$^3$)\\
      \hline \\
  IK~Tau     & -4.8~--~-3.5& -3.0~--~-1.5& -1.2~--~0.5 & 0.7~--~2.4\\
  RT~Vir     & -4.2~--~-3.8& -2.4~--~-1.3& -0.7~--~1.2 & 1.1~--~3.7\\
  IRC$+$60370& -5.1~--~-3.2& -3.4~--~-0.7& -1.7~--~1.7 & 0.0~--~4.2\\
      \hline \\
      U (J/m$^3$) & \h2o & Star &\\
      \hline \\
      $nKT$           & $\sim$-6.2 & $\sim$0.5&&\\
      $\rho V_{exp}^2$ & $\sim$-5.1 & $\sim$0.5&&\\
      \hline
    \end{tabular}
    \label{tabBenergy}
  \end{center}
  {In the upper part of the table, from Cols. 1 to 4: the source name 
(Source), the log of the magnetic energy density ($U=B^2/2\mu_0$) 
in the \h2o maser region ($U_{H_2O}$), in the stellar surface assuming $\alpha$ 
equal to one ($U_{star}$ ($\alpha$=1)), in the stellar surface assuming 
$\alpha$ equal to two ($U_{star}$ ($\alpha$=2)), and in the stellar surface 
assuming $\alpha$ equal to three ($U_{star}$ ($\alpha$=3)). In the lower 
part of the table, we show the log of typical values for kinetic and 
thermal energy densities (Col.~1; Energy density) in the \h2o maser region 
(Col. 2; \h2o) and on the stellar surface (Col. 3; Star).}
\end{table}

\section{Conclusions} 
\label{conclusions}

We observed four AGB stars in order to detect \h2o maser in full polarization 
at VLBI resolution. We did not detect any maser emission toward AP~Lyn. Toward 
IK~Tau, RT~Vir, and IRC$+$60370 we detected 85, 91, and 62 features, 
respectively.

A structured spatial distribution of maser velocities was observed toward 
IK~Tau. This behavior has already been reported by \cite{bains03} and an 
equatorial density enhancement model was proposed. A similar signature, but 
less pronounced, was observed toward RT~Vir, but with opposite 
velocity-position pattern to those reported by \cite{bains03}.

We used the shell-fitting method to infer the projected position of the star 
relative to the observed masers. With the stellar position determined, we 
produced a plot of the angular offset of the masers relative to the stellar 
position versus the maser velocities. We fitted parabolas in these plots to 
determine the actual distance of the \h2o maser regions from the central 
stars. We concluded that the \h2o masers we observed are located between 10.1 
and 29.2~AU from IK~Tau, 2.4 and 18.0~AU from RT~Vir, and 5.5 and 53.0~AU from 
IRC$+$60370.

Linear polarization was observed in 18 features, nine around RT~Vir 
and nine around IRC$+$60370. Circular polarization was found in 11 features, 
three around IK~Tau, three around RT~Vir, and five around IRC$+$60370. 
From a model fit of the Stokes V spectra of the features with statistically 
significant circular polarization detection, we estimated the magnetic field 
strength along the line of sight needed to generate the observed S-shape 
profile. The resulting projected magnetic field strengths lie between 
47$\pm$3~mG and 331$\pm$82~mG. With our polarization results, we more 
than doubled the number of AGB stars around which the magnetic field has 
been detected in the \h2o maser region.

Combining our results with published results for the magnetic field 
measurements in the SiO maser regions, it is not yet possible to determine the 
magnetic field dependence on the radial distance $R$ to the star. For IK~Tau, 
either a dependence $B \propto$~$R^{-2}$ or $B \propto$~$R^{-3}$ seems 
qualitatively more likely, but $B \propto$~$R^{-1}$ is not ruled out. The 
results we found in the literature for RT~Vir and IRC$+$60370 are not 
sufficient to draw stronger conclusions.

The results we obtained for the magnetic field strength along the line 
of sight were extrapolated to the stellar surface of the observed sources, 
assuming $B \propto$~$R^{-1}$, $B \propto$~$R^{-2}$, and $B \propto$~$R^{-3}$. 
In the first case, the projected field strength on the AGB star surface 
($B_{star}$) should be between 0.5~G and 2.7~G for IK~Tau, 1.0~G and 3.5~G 
for RT~Vir, and 0.3~G and 6.9~G for IRC$+$60370. If $B \propto$~$R^{-2}$ is 
assumed, then $B_{star}$ was extrapolated to be between 4.2~G and 26.8~G for 
IK~Tau, 7.3~G and 64.9~G for RT~Vir, and 2.2~G and $\sim$115~G for IRC$+$60370. 
If $B \propto$~$R^{-3}$, then $B_{star}$ was found to be between 33.8~G 
and $\sim$265~G for IK~Tau, 54.9~G and $\sim$1185~G for RT~Vir, and 
15.8~G and $\sim$1945~G for IRC$+$60370. 

Finally, we compared the magnetic energy density we observed with the typical 
thermal and kinematic energy density around evolved stars. While the dominant 
energy density on the surface of the star is still inconclusive, we conclude 
that, in the \h2o maser region, the magnetic energy density dominates the 
thermal and kinematic energy density. Therefore, the magnetic fields cannot be 
ignored as one of the important agents in shaping planetary nebulae.

 
\begin{acknowledgements} 
The authors would like to thank Anita Richards, the referee of this 
paper, for her comments that helped to significantly improve the article. This 
research was supported by the Deutscher Akademischer Austausch Dienst (DAAD) 
and the Deutsche Forschungsgemeinschaft (DFG; through the Emmy Noether 
Research grant VL 61/3-1).
\end{acknowledgements} 
 

\longtab{7}{
\begin{longtable}{lcrrrrrcccccc}
\caption{\label{results} Properties of the observed masers. Respectively, from 
Cols. 1 to 10: the source in which the maser was observed (Source), the 
maser identification (feature), projected offset on the plane of sky in 
direction of right ascension ($\alpha_{off}$), offset in declination 
($\delta_{off}$), peak intensity (Peak Int), intensity flux (Int Flux), 
velocity of the peak (V$_{peak}$), P$_V$, magnetic field strength 
(B $cos \theta$), and if linear polarization was detected or not (LinPol). 
$\alpha_{off}$ and $\delta_{off}$ are with respect to the reference feature.}\\
\hline\hline
Source &Feature & $\alpha_{off}$ & $\delta_{off}$ & Peak Int & Int Flux & V$_{peak}$ & P$_L$ &EVPA     & P$_V$           &B$_{||}$ \\
       &        & (mas)         & (mas)         & (Jy/Beam)& (Jy)      & (km/s)    &($\%$) &($^\circ$)&($\times 10^{-3}$)& (mG) \\
\hline
\endfirsthead
\caption{continued.}\\
\hline\hline
Source&Feature&$\alpha_{off}$&$\delta_{off}$&Peak Int&Int Flux&V$_{peak}$& P$_L$ & EVPA & P$_V$ &B$_{||[Gauss]}$\\ 
      &       & (mas)       & (mas)       &(Jy/Beam)& (Jy)   & (km/s)  &($\%$)&($^\circ$)&($\times 10^{-3}$)&(mG)\\
\hline
\endhead
\hline
\endfoot

\hline
AP~Lyn      & --  & --      & --       & --   & --   & --    & -- & --     & -- & --\\  
\hline
IK~Tau      &IK.01 &   7.5&   6.6&  0.20&  0.35&   42.6&--  &--      &--  & -- \\
            &IK.02 &   3.2&   9.3&  0.08&  0.15&   42.5&--  &--      &--  & -- \\
            &IK.03 &  52.3&--37.0&  0.06&  0.18&   42.2&--  &--      &--  & -- \\
            &IK.04 &  52.6&--38.6&  0.33&  0.54&   42.1&--  &--      &--  & -- \\
            &IK.05 &   3.6& --6.6&  0.20&  0.41&   42.1&--  &--      &--  & -- \\
            &IK.06 &   4.7& --6.4&  0.17&  0.54&   42.1&--  &--      &--  & -- \\
            &IK.07 &   7.1&   3.9&  0.04&  0.09&   42.1&--  &--      &--  & -- \\
            &IK.08 &   1.3&   6.6&  0.03&  0.05&   42.1&--  &--      &--  & -- \\
            &IK.09 &   3.0& --6.6&  0.24&  0.52&   42.0&--  &--      &--  & -- \\
            &IK.10 &  53.7&--38.3&  0.30&  0.53&   41.9&--  &--      &--  & -- \\
            &IK.11 &   1.8& --4.5&  0.08&  0.13&   41.9&--  &--      &--  & -- \\
            &IK.12 & --3.1& --4.4&  0.22&  0.50&   41.9&--  &--      &--  & -- \\
            &IK.13 &  55.0&--38.5&  0.27&  0.87&   41.8&--  &--      &--  & -- \\
            &IK.14 &  51.3&--40.6&  0.21&  0.37&   41.7&--  &--      &--  & -- \\
            &IK.15 &  31.7&--41.4&  0.14&  0.25&   41.6&--  &--      &--  & -- \\
            &IK.16 &  53.8&--43.2&  0.12&  0.20&   41.4&--  &--      &--  & -- \\
            &IK.17 &  50.5&--48.1&  0.18&  0.30&   41.0&--  &--      &--  & -- \\
            &IK.18 &  53.3&--44.5&  0.07&  0.10&   40.7&--  &--      &--  & -- \\
            &IK.19 &  61.1&--26.4&  0.15&  0.31&   40.7&--  &--      &--  & -- \\
            &IK.20 & --0.0&   0.0&  4.93& 10.22&   40.5&--  &--      &10.08$\pm$1.03&$-$147$\pm$15\\
            &IK.21 &  63.3&--31.1&  0.04&  0.06&   40.2&--  &--      &--  & -- \\
            &IK.22 &  66.0&--32.1&  0.07&  0.17&   40.0&--  &--      &--  & -- \\
            &IK.23 &  66.4&--32.9&  0.06&  0.14&   39.9&--  &--      &--  & -- \\
            &IK.24 &  67.2&--34.1&  0.04&  0.07&   39.6&--  &--      &--  & -- \\
            &IK.25 &--16.1&  22.8&  0.06&  0.14&   38.9&--  &--      &--  & -- \\
            &IK.26 &   0.5&   9.5&  0.21&  0.42&   38.7&--  &--      &--  & -- \\
            &IK.27 & --0.4&   9.8&  0.11&  0.26&   38.7&--  &--      &--  & -- \\
            &IK.28 &  73.5&--16.0&  0.11&  0.28&   38.6&--  &--      &--  & -- \\
            &IK.29 &  73.0&--17.0&  0.17&  0.92&   38.4&--  &--      &--  & -- \\
            &IK.30 &--35.8&  45.1&  0.61&  1.06&   31.7&--  &--      &--  & -- \\
            &IK.31 &--34.5&  41.9&  0.34&  0.77&   31.6&--  &--      &--  & -- \\
            &IK.32 &--36.8&  46.9&  0.75&  1.44&   31.4&--  &--      &--  & -- \\
            &IK.33 &  13.9&--40.6&  0.07&  0.18&   31.4&--  &--      &--  & -- \\
            &IK.34 &--33.3&  40.3&  1.33&  2.54&   31.0&--  &--      &--  & -- \\
            &IK.35 &  13.9&--42.5&  0.06&  0.13&   31.0&--  &--      &--  & -- \\
            &IK.36 & --5.0&  13.5&  0.06&  0.10&   30.7&--  &--      &--  & -- \\
            &IK.37 & --6.9&  13.3&  0.19&  0.59&   30.3&--  &--      &--  & -- \\
            &IK.38 &--34.1&  44.7&  0.08&  0.15&   30.3&--  &--      &--  & -- \\
            &IK.39 & --7.9&  12.9&  0.12&  0.45&   30.2&--  &--      &--  & -- \\
            &IK.40 & --6.7&   7.6&  0.07&  0.15&   30.2&--  &--      &--  & -- \\
            &IK.41 & --8.9&  12.5&  0.11&  0.24&   30.0&--  &--      &--  & -- \\
            &IK.42 & --8.1&   8.1&  0.17&  0.57&   29.9&--  &--      &--  & -- \\
            &IK.43 &   1.0&   2.3&  0.12&  0.31&   29.9&--  &--      &--  & -- \\
            &IK.44 & --0.8&   2.0&  1.11&  3.12&   29.8&--  &--      &--  & -- \\
            &IK.45 & --9.2&   8.0&  0.17&  0.49&   29.8&--  &--      &--  & -- \\
            &IK.46 & --5.8&   5.6&  0.07&  0.12&   29.5&--  &--      &--  & -- \\
            &IK.47 & --9.0&   9.5&  0.13&  0.29&   29.5&--  &--      &--  & -- \\
            &IK.48 & --9.8&   9.4&  0.10&  0.22&   29.5&--  &--      &--  & -- \\
            &IK.49 &--18.0&   7.0&  0.06&  0.10&   29.4&--  &--      &--  & -- \\
            &IK.50 &--10.5&  10.4&  0.08&  0.19&   29.4&--  &--      &--  & -- \\
            &IK.51 & --2.9&   1.4&  0.12&  0.28&   29.3&--  &--      &--  & -- \\
            &IK.52 & --6.7&  26.9&  0.06&  0.11&   29.3&--  &--      &--  & -- \\
            &IK.53 & --3.7&   1.3&  0.13&  0.26&   29.2&--  &--      &--  & -- \\
            &IK.54 &--10.6&  10.3&  0.07&  0.18&   29.2&--  &--      &--  & -- \\
            &IK.55 &   0.8& --1.6&  0.05&  0.10&   28.8&--  &--      &--  & -- \\
            &IK.56 & --0.0& --2.2&  0.09&  0.51&   28.7&--  &--      &--  & -- \\
            &IK.57 &--24.3&   5.0&  0.09&  0.20&   28.6&--  &--      &--  & -- \\
            &IK.58 & --0.5& --1.3&  0.15&  0.28&   28.2&--  &--      &--  & -- \\
            &IK.59 &  19.7&  59.5&  0.08&  0.15&   28.1&--  &--      &--  & -- \\
            &IK.60 &--13.1&   3.6&  0.06&  0.11&   28.1&--  &--      &--  & -- \\
            &IK.61 &  19.1&  63.5&  0.60&  1.11&   28.0&--  &--      &--  & -- \\
            &IK.62 &--12.0&   2.2&  0.10&  0.21&   28.0&--  &--      &--  & -- \\
            &IK.63 &  18.1&  62.6&  0.14&  0.21&   27.9&--  &--      &--  & -- \\
IK~Tau      &IK.64 &--27.6&  46.0&  0.08&  0.14&   27.8&--  &--      &--  & -- \\
            &IK.65 & --6.7& --2.6&  0.45&  1.14&   27.7&--  &--      &--  & -- \\
            &IK.66 &--10.0& --2.6&  0.29&  0.88&   27.7&--  &--      &--  & -- \\
            &IK.67 & --8.6& --2.1&  0.20&  0.56&   27.7&--  &--      &--  & -- \\
            &IK.68 & --7.7& --2.4&  0.18&  0.60&   27.6&--  &--      &--  & -- \\
            &IK.69 &  16.4&  74.5&  2.97&  4.28&   27.4&--  &--      &5.48$\pm$1.78&$-$96$\pm$31\\
            &IK.70 &--43.0&--25.1&  0.81&  1.66&   27.2&--  &--      &--  & -- \\
            &IK.71 &--44.0&--24.8&  0.32&  0.88&   27.2&--  &--      &--  & -- \\
            &IK.72 & --6.0& --2.2&  0.16&  0.35&   27.2&--  &--      &--  & -- \\
            &IK.73 &--49.9&--26.3&  0.23&  0.53&   27.0&--  &--      &--  & -- \\
            &IK.74 &  10.0&  64.1&  0.15&  0.19&   26.8&--  &--      &--  & -- \\
            &IK.75 &  19.7&  73.3&  0.17&  0.27&   26.6&--  &--      &--  & -- \\
            &IK.76 &--34.4&--28.6&  0.12&  0.17&   26.4&--  &--      &--  & -- \\
            &IK.77 &  20.9&  34.5&  0.08&  0.13&   25.9&--  &--      &--  & -- \\
            &IK.78 &  55.8&  14.6&  2.37&  4.67&   25.7&--  &--      &--  & -- \\
            &IK.79 &   6.9&--51.7&  0.07&  0.15&   25.6&--  &--      &--  & -- \\
            &IK.80 &  57.2&  15.1&  0.08&  0.14&   25.3&--  &--      &--  & -- \\
            &IK.81 &  55.1&  15.3&  0.78&  1.96&   25.3&--  &--      &--  & -- \\
            &IK.82 &  53.8&  15.6&  0.30&  0.88&   25.1&--  &--      &--  & -- \\
            &IK.83 &  50.8&  15.0&  0.86&  2.50&   25.0&--  &--      &--  & -- \\
            &IK.84 &  51.2&  15.7&  0.90&  2.18&   25.0&--  &--      &12.15$\pm$3.14&$+$215$\pm$56\\
            &IK.85 &  55.1&  15.5&  0.59&  1.29&   24.8&--  &--      &--  & -- \\
\hline                                                           
RT~Vir      &RT.01 &--19.3& --1.4&  0.14&  0.16&   21.8&--  &--      &--  & --\\      
            &RT.02 &  45.9&--67.7&  0.07&  0.16&   21.8&--  &--      &--  & --\\   
            &RT.03 &  47.6&--64.9&  0.09&  0.13&   21.6&--  &--      &--  & --\\   
            &RT.04 &  13.4&   8.5&  0.03&  0.04&   21.3&--  &--      &--  & --\\   
            &RT.05 &  41.8&--73.8&  0.05&  0.10&   21.2&--  &--      &--  & --\\   
            &RT.06 &--15.6&--50.1&  0.06&  0.07&   21.1&--  &--      &--  & --\\   
            &RT.07 &   8.8&  11.9&  0.06&  0.08&   21.0&--  &--      &--  & --\\   
            &RT.08 &--12.7&--54.0&  0.03&  0.04&   21.0&--  &--      &--  & --\\   
            &RT.09 & --8.1&--22.9&  0.29&  0.42&   20.9&--  &--      &--  & --\\   
            &RT.10 & --7.2&--21.5&  0.22&  0.30&   20.8&--  &--      &--  & --\\   
            &RT.11 &--28.0&   2.2&  0.04&  0.06&   20.8&--  &--      &--  & --\\   
            &RT.12 & --9.2&--21.6&  0.12&  0.21&   20.5&--  &--      &--  & --\\   
            &RT.13 &--10.4&--22.4&  0.26&  0.37&   20.4&--  &--      &--  & --\\   
            &RT.14 &--12.4&--77.2&  0.04&  0.09&   20.4&--  &--      &--  & --\\     
            &RT.15 &--14.5&--34.0&  0.64&  1.44&   20.1&--  &--      &--  & --\\     
            &RT.16 &--36.4&--60.4&  0.07&  0.09&   20.1&--  &--      &--  & --\\     
            &RT.17 &--10.5&--22.4&  0.21&  0.29&   19.9&--  &--      &--  & --\\    
            &RT.18 &  52.6&--72.5&  0.29&  0.40&   18.1&--  &--      &--  & --\\
            &RT.19 &  52.8&--55.5&  0.33&  0.61&   18.1&--  &--      &--  & --\\
            &RT.20 &  49.1&--51.8&  0.74&  1.13&   18.1&--  &--      &--  & --\\
            &RT.21 &  53.1&--49.9&  4.65&  5.42&   18.1&--  &--      &--  & --\\
            &RT.22 &--26.1&--15.8&  0.25&  0.40&   18.1&--  &--      &--  & --\\
            &RT.23 &--25.3&   6.0&  3.13&  5.03&   18.1&--  &--      &--  & --\\
            &RT.24 &--23.5&   6.0&  0.77&  0.99&   18.0&--  &--      &--  & --\\
            &RT.25 &  53.5&  60.1&  0.18&  0.34&   18.0&--  &--      &--  & --\\
            &RT.26 &--16.7& --3.1&  0.90&  0.97&   17.7&--  &--      &--  & --\\
            &RT.27 &--29.8&  27.2&  2.34&  3.73&   17.6&--  &--      &--  & --\\
            &RT.28 &--32.4&  27.1&  0.62&  0.66&   17.5&--  &--      &--  & --\\
            &RT.29 &--37.1&--50.0&  0.56&  0.72&   17.1&--  &--      &--  & --\\
            &RT.30 &--31.4&  25.6&  7.83&  9.24&   17.0&--  &--      &--  & --\\
            &RT.31 &--35.8&  25.9& 20.96& 25.07&   17.0&0.22$\pm$0.02&$-$46$\pm$5&--  & --\\
            &RT.32 &--35.0&  28.0&  2.83&  3.55&   17.0&--  &--      &--  & --\\
            &RT.33 & --9.6&--28.8&  0.60&  0.90&   17.0&--  &--      &--  & --\\
            &RT.34 &--17.0& --5.4&  1.21&  1.38&   16.7&1.41$\pm$0.04&$-$59$\pm$3&--  & --\\
            &RT.35 &--30.8&  28.4&  0.79&  0.86&   16.6&--  &--      &--  & --\\
            &RT.36 &--32.3&  25.1&  1.32&  3.13&   16.4&--  &--      &--  & --\\
            &RT.37 &  39.8&--41.5&  0.32&  0.51&   16.3&--  &--      &--  & --\\
            &RT.38 &--21.8&  10.7&  0.28&  0.35&   16.3&--  &--      &--  & --\\
            &RT.39 &  29.7&--75.6&  0.48&  0.75&   15.9&--  &--      &--  & --\\
            &RT.40 &  90.7&  18.0&  0.68&  1.08&   15.6&--  &--      &--  & --\\
            &RT.41 &  79.5&  36.2&  0.48&  0.89&   15.6&--  &--      &--  & --\\
            &RT.42 &  86.2&  89.1&  0.44&  0.54&   15.6&--  &--      &--  & --\\
            &RT.43 &--14.3& --7.5&  0.59&  1.58&   15.4&--  &--      &--  & --\\
            &RT.44 &  87.8& --2.1&  0.64&  1.21&   15.4&--  &--      &--  & --\\
RT~Vir      &RT.45 &--15.4& --7.4&  1.15&  1.62&   15.3&--  &--      &--  & --\\
            &RT.46 &  23.6&--51.0&  0.22&  0.48&   15.2&--  &--      &--  & --\\
            &RT.47 &  44.3&--44.3&  0.28&  0.36&   15.2&--  &--      &--  & --\\
            &RT.48 &  31.7&--82.4&  0.39&  0.59&   14.9&--  &--      &--  & --\\
            &RT.49 & --6.7& --8.0&  0.26&  0.56&   14.8&--  &--      &--  & --\\
            &RT.50 &--31.3&  27.6&  0.12&  0.12&   14.8&--  &--      &--  & --\\
            &RT.51 &--39.3&  20.4&  0.10&  0.13&   14.8&--  &--      &--  & --\\
            &RT.52 &  33.9&--90.5&  0.25&  0.36&   14.6&--  &--      &--  & --\\
            &RT.53 &  31.8&--36.2&  0.32&  0.35&   14.5&--  &--      &--  & --\\
            &RT.54 &  58.5&--16.8&  1.20&  1.33&   14.3&--  &--      &--  & --\\
            &RT.55 &  58.2&--50.3&  0.08&  0.11&   14.3&--  &--      &--  & --\\
            &RT.56 &--22.9&--22.1&  0.08&  0.18&   14.3&--  &--      &--  & --\\
            &RT.57 &--20.9&--20.9&  0.23&  0.41&   14.2&--  &--      &--  & --\\
            &RT.58 &   2.8&--18.3&  0.13&  0.15&   14.0&--  &--      &--  & --\\
            &RT.59 &  29.4&--15.8&  0.34&  0.46&   13.9&--  &--      &--  & --\\
            &RT.60 &   3.6&   4.2&  0.58&  0.97&   13.8&--  &--      &--  & --\\
            &RT.61 &  30.6&--16.4&  0.15&  0.20&   13.8&--  &--      &--  & --\\
            &RT.62 &  29.9&--14.0&  0.35&  0.39&   13.7&--  &--      &--  & --\\
            &RT.63 &   2.2&   1.8&  0.42&  0.48&   13.7&--  &--      &--  & --\\
            &RT.64 &   2.4&  19.0&  0.36&  0.40&   13.7&--  &--      &--  & --\\
            &RT.65 & --1.9& --1.0&  0.57&  0.71&   13.6&--  &--      &--  & --\\
            &RT.66 &   5.2&--54.0&  0.36&  0.49&   13.4&--  &--      &--  & --\\
            &RT.67 & --1.0& --1.8&  2.08&  2.71&   13.1&0.56$\pm$0.26&$-$44$\pm$16&--  & --\\
            &RT.68 &   0.0&   0.0& 54.53& 63.13&   12.9&0.48$\pm$0.19&$-$38$\pm$12&--  & --\\
            &RT.69 &   8.0&--85.1&  1.30&  1.60&   12.9&--  &--      &--  & --\\
            &RT.70 &  30.9&--10.6&  7.92&  8.45&   12.9&1.12$\pm$0.28&$+$49$\pm$26&10.50$\pm$0.86&$-$143$\pm$12$^e$\\
            &RT.71 & --0.3&  55.2&  0.93&  1.29&   12.9&--  &--      &--  & --\\
            &RT.72 & --7.5&  12.0&  4.23&  5.28&   12.4&0.38$\pm$0.29&$-$51$\pm$41&--  & --\\
            &RT.73 & --0.9& --1.6&  6.94&  7.53&   12.0&0.49$\pm$0.36&$-$45$\pm$24&--  & --\\
            &RT.74 &  57.2&--64.0&  1.19&  1.72&   11.8&--  &--      &--  & --\\
            &RT.75 &  31.0& --9.1& 39.52& 48.67&   11.7&0.63$\pm$0.12&$+$64$\pm$7&6.05$\pm$0.19&$-$188$\pm$6\\
            &RT.76 &  28.3& --6.4&  4.08&  5.32&   10.7&--  &--      &--  & --\\
            &RT.77 &  25.2&--61.4&  0.31&  0.69&   10.7&--  &--      &--  &-- \\
            &RT.78 &  24.0&--60.4&  0.32&  0.67&   10.7&--  &--      &--  &-- \\
            &RT.79 &  28.0&--27.4&  0.29&  0.41&   10.6&--  &--      &--  &-- \\
            &RT.80 &--34.5&   8.4&  0.21&  0.64&   10.6&--  &--      &--  &-- \\
            &RT.81 &  26.8&--41.0&  0.06&  0.08&   10.2&--  &--      &--  &-- \\
            &RT.82 &  29.3& --1.6&  0.11&  0.13&   9.9 &--  &--      &--  &-- \\
            &RT.83 &  29.9&   4.0&  0.45&  0.49&   9.7 &--  &--      &--  &-- \\
            &RT.84 &  28.9&  19.1&  0.14&  0.14&   9.7 &--  &--      &--  &-- \\
            &RT.85 &  21.7&   7.7&  2.72&  3.09&   9.6 &--  &--      &--  &-- \\
            &RT.86 &  23.8&   2.1&  0.53&  0.88&   9.4 &--  &--      &--  &-- \\
            &RT.87 &  26.2&   0.4&  0.42&  0.51&   9.4 &--  &--      &--  &-- \\
            &RT.88 &  21.3&   3.1&  3.02&  3.57&   9.3 &--  &--      &--  & --\\
            &RT.89 &  21.8&   5.2&  0.64&  0.80&   9.2 &--  &--      &--  &-- \\
            &RT.90 &  24.7&   3.8& 23.27& 28.03&   8.9 &0.11$\pm$0.01&$+$38$\pm$3&1.80$\pm$0.48&$-$84 \& $+$63$^b$\\
            &RT.91 &  50.8&--51.2&  0.52&  0.61&   8.8 &--  &--      &--  &-- \\
\hline                                                           
IRC$+$60370 &IRC.01& --6.5& 14.5&  0.12&  0.13&$-$39.7&--  &--      &--  & --\\
            &IRC.02& --5.7& 16.1&  0.11&  0.12&$-$40.0&--  &--      &--  & --\\
            &IRC.03&  33.2&-35.7&  0.89&  1.10&$-$40.2&--  &--      &--  & --\\
            &IRC.04& --8.4& 13.7&  0.18&  0.29&$-$40.5&--  &--      &--  & --\\
            &IRC.05&   4.7&-10.0&  0.63&  0.75&$-$40.6&--  &--      &--  & --\\
            &IRC.06& --4.7& 15.8&  0.20&  0.24&$-$40.8&--  &--      &--  & --\\
            &IRC.07&  14.3&-10.3&  0.17&  0.22&$-$40.9&--  &--      &--  & --\\
            &IRC.08&  12.4& -9.9&  0.10&  0.16&$-$41.3&--  &--      &--  & --\\
            &IRC.09& --3.8& 16.3&  0.05&  0.07&$-$41.3&--  &--      &--  & --\\
            &IRC.10&   5.0& -9.4&  0.16&  0.22&$-$41.4&--  &--      &--  & --\\
            &IRC.11&  17.2& -8.8&  0.05&  0.06&$-$43.6&--  &--      &--  & --\\
            &IRC.12&   5.6& 13.2&  0.77&  1.00&$-$43.7&--  &--      &--  & --\\
            &IRC.13&  34.6&-24.3&  0.08&  0.08&$-$43.7&--  &--      &--  & --\\
            &IRC.14&  28.1& 13.0&  0.08&  0.12&$-$43.8&--  &--      &--  & --\\
            &IRC.15&--17.2& -6.0&  0.11&  0.14&$-$44.2&--  &--      &--  & --\\
            &IRC.16&  18.6&-10.5&  0.05&  0.05&$-$44.2&--  &--      &--  & --\\
            &IRC.17&  27.9& 12.3&  0.09&  0.16&$-$44.4&--  &--      &--  & --\\
            &IRC.18&  35.4&-23.4&  0.11&  0.11&$-$44.4&--  &--      &--  & --\\
            &IRC.19&   6.5& 13.0&  1.03&  1.24&$-$44.5&--  &--      &--  & --\\
IRC$+$60370 &IRC.20&   3.9& 13.7&  0.10&  0.14&$-$44.8&--  &--      &--  & --\\
            &IRC.21&  15.6& -7.2&  0.17&  0.21&$-$45.2&--  &--      &--  & --\\
            &IRC.22&  33.2&-43.1&  0.06&  0.06&$-$45.2&--  &--      &--  & --\\
            &IRC.23&  29.6&  9.9&  0.13&  0.21&$-$45.3&--  &--      &--  & --\\
            &IRC.24&   2.8& 13.1&  0.08&  0.11&$-$45.3&--  &--      &--  & --\\
            &IRC.25&   4.8& 11.0&  0.10&  0.14&$-$45.6&--  &--      &--  & --\\
            &IRC.26&   4.0&  9.6&  0.11&  0.31&$-$45.8&--  &--      &--  & --\\
            &IRC.27&   3.7&  8.4&  2.04&  3.29&$-$47.0&--  &--      &--  & --\\
            &IRC.28&   4.5& 10.0&  0.05&  0.08&$-$47.4&--  &--      &--  & --\\
            &IRC.29& --7.6&  4.5&  0.15&  0.19&$-$47.9&--  &--      &--  & --\\
            &IRC.30& --8.9&  4.4&  0.20&  0.35&$-$48.0&--  &--      &--  & --\\
            &IRC.31& --0.8&  5.9&  0.10&  0.10&$-$48.0&--  &--      &--  & --\\
            &IRC.32& --5.6&  4.4&  0.08&  0.15&$-$48.0&--  &--      &--  & --\\
            &IRC.33&   2.8&  8.0&  0.05&  0.06&$-$48.0&--  &--      &--  & --\\
            &IRC.34&  27.8&  7.2&  1.14&  3.31&$-$48.9&1.51$\pm$0.10&$-$132$\pm$2&--  & --\\
            &IRC.35&  28.3&  6.8&  6.29& 11.16&$-$49.1&0.61$\pm$0.07&$-$67$\pm$7&--  & -- \\
            &IRC.36&  28.3& -5.7&  1.29&  1.47&$-$49.3&--  &--      &--  & -- \\
            &IRC.37&   3.6&  2.8&  1.19&  1.52&$-$49.7&--  &--      &--  & -- \\
            &IRC.38&  27.2&  6.5&  3.03&  4.35&$-$49.8&0.58$\pm$0.03&$-$74$\pm$2&--  & --\\
            &IRC.39&   5.6&  5.8&  0.17&  0.24&$-$49.9&--  &--      &--  & --\\
            &IRC.40&  30.6&  5.8&  0.22&  0.34&$-$50.1&--  &--      &--  & --\\
            &IRC.41& --1.1& -0.4&  1.96&  2.54&$-$50.3&1.58$\pm$0.30& $-$39$\pm$7&--  & --\\
            &IRC.42& --0.9&  0.2& 11.91& 20.83&$-$51.4&0.65$\pm$0.30&$-$58$\pm$16&--  & --\\
            &IRC.43&   4.4& -8.6&  1.65&  2.05&$-$51.8&--  &--      &--  & -- \\ 
            &IRC.44&   0.0&  0.0& 51.23& 67.03&$-$52.0&0.57$\pm$0.02&$-$77$\pm$2&2.10$\pm$0.13&$+$47$\pm$3$^e$\\  
            &IRC.45& --0.6&  2.0&  5.21&  8.88&$-$52.3&0.45$\pm$0.07&$-$-93$\pm$5&10.71$\pm$1.21&$+$266$\pm$30$^e$\\
            &IRC.46& --1.3&  2.0&  1.68&  3.06&$-$52.7&--  &--      &--  & -- \\
            &IRC.47& --2.0&  0.9&  1.65&  2.24&$-$53.0&--  &--      &15.52$\pm$4.05&$+$331$\pm$82$^e$\\
            &IRC.48& --0.8&  0.6& 11.26& 15.17&$-$53.3&0.46$\pm$0.18&$-$75$\pm$14&10.38$\pm$0.61&$+$273$\pm$18$^e$\\  
            &IRC.49&  23.2& 12.6&  0.34&  0.48&$-$53.8&--  &--      &--  & --\\
            &IRC.50&  23.3& 12.8&  0.33&  0.51&$-$54.0&--  &--      &--  & --\\
            &IRC.51& --1.6&  2.8&  9.08&  9.43&$-$54.2&0.19$\pm$0.02&$-$97$\pm$6&--  & --\\ 
            &IRC.52&--20.0& 13.0&  0.14&  0.26&$-$54.6&--  &--      &--  & --\\
            &IRC.53&  16.7& 20.4&  0.06&  0.07&$-$55.8&--  &--      &--  & --\\
            &IRC.54&  29.6&  3.9&  0.40&  0.49&$-$55.9&--  &--      &--  & --\\
            &IRC.55&  29.2&  4.0&  0.09&  0.14&$-$56.8&--  &--      &--  & --\\
            &IRC.56&  26.4& 21.2&  0.03&  0.04&$-$57.4&--  &--      &--  & --\\
            &IRC.57&  28.8&  4.0&  0.23&  0.53&$-$57.6&--  &--      &--  & --\\
            &IRC.58&  28.5&  4.6&  4.89&  5.48&$-$58.3&--  &--      &8.34$\pm$1.40&$-$130$\pm$22\\
            &IRC.59&  28.7&  3.7&  0.20&  0.29&$-$59.0&--  &--      &--  & --\\
            &IRC.60&  14.6& -4.8&  0.53&  0.57&$-$59.5&--  &--      &--  & --\\
            &IRC.61&  14.4& -4.5&  0.76&  0.81&$-$60.6&--  &--      &--  & --\\
            &IRC.62&   7.0& 14.4&  0.07&  0.09&$-$63.2&--  &--      &--  & --\\
      \hline                                 
                               
\multicolumn{7}{l}{$^e$ Edge/higher noise effects}\\
\multicolumn{7}{l}{$^b$ Blended feature}\\
\end{longtable}}               
                               

\begin{thebibliography}{99}
\bibitem[\protect\citeauthoryear{Amiri et al.}{2011}]{amiri11}
Amiri,N., Vlemmings, W., van Langevelde, H. J. 2011, A\&A, 532, 149 
\bibitem[\protect\citeauthoryear{Bains et al.}{2003}]{bains03}
Bains, I., Cohen, R. J., Louridas, A., Richards, A. M. S., Rosa-González, D., 
Yates, J. A. 2003, MNRAS, 342, 8
\bibitem[\protect\citeauthoryear{Balick \& Frank}{2002}]{balick02} 
Balick, B., Frank, A. 2002 , ARA\&A, 40, 439
\bibitem[\protect\citeauthoryear{Boboltz \& Diamond}{2005}]{boboltz05}
Boboltz, D. A., Diamond, P. J. 2005, ApJ, 625, 978
\bibitem[\protect\citeauthoryear{Bowers et al.}{1989}]{bowers89}
Bowers, P. F., Johnston, K. J., de Vegt, C. 1989, ApJ, 340, 479
\bibitem[\protect\citeauthoryear{Castro-Carrizo}{2010}]{castro10}
Castro-Carrizo, A., Quintana-Cacaci, G., Neri, R., Bujarrabal, V., Sch\"oier, 
F. L., Winters, J. M., Olofsson, H., Lindqvist, M., Alcolea, J., Lucas, R., 
Grewing, M. 2010, A\&A, 523, 59
\bibitem[\protect\citeauthoryear{Cohen}{1987}]{cohen87}
Cohen, R. J., 1987, IAUS, 122, 229
\bibitem[\protect\citeauthoryear{Colomer et al.}{2000}]{colomer00} 
Colomer, F., Reid, M. J., Menten, K. M., Bujarrabal, V. 2000, A\&A, 355, 979
\bibitem[\protect\citeauthoryear{Dennis et al.}{2009}]{dennis09} 
Dennis, T. J., Frank, A., Blackman, E. G., De Marco, O., Balick, B., Mitran, 
S. 2009, ApJ, 707, 1485
\bibitem[\protect\citeauthoryear{Elitzur}{1992}]{elitzur92} 
Elitzur, M. 1992, ARA\&A, 30, 75
\bibitem[\protect\citeauthoryear{Frank et al.}{2007}]{frank07} 
Frank, A., De Marco, O., Blackman, E., Balick, B. 2007, unpublished, 
arXiv:0712.2004
\bibitem[\protect\citeauthoryear{Garc\'ia-Segura et al.}{1999}]{garcia99} 
Garc\'ia-Segura, G., Langer, N., R\'o\.zyczka, M., Franco, J. 1999, ApJ, 517, 
767
\bibitem[\protect\citeauthoryear{Garc\'ia-Segura et al.}{2005}]{garcia05} 
Garc\'ia-Segura, G., L\'opez, J. A., Franco, J. 2005, ApJ, 618, 919
\bibitem[\protect\citeauthoryear{Garc\'ia-D\'iaz et al.}{2008}]{garcia08} 
Garc\'ia-D\'iaz, M. T., L\'opez, J. A., Richer, M. G., Steffen, W. 2008, ApJ, 
676, 402
\bibitem[\protect\citeauthoryear{Goldreich et al.}{1973}]{goldreich73}
Goldreich, P., Keeley, D. A., Kwan, J. Y. 1973, ApJ, 179, 111
\bibitem[\protect\citeauthoryear{Herpin et al.}{2006}]{herpin06}
Herpin, F., Baudry, A., Thum, C., Morris, D., Wiesemeyer, H. 2006, A\&A, 450, 
667
\bibitem[\protect\citeauthoryear{Hipparcos}{1997}]{hipparcos97}
Hipparcos Catalogue, 1997, ESA SP-1200, CDS-VizieR 
(http://vizier.u-strasbg.fr/viz-bin/Cat?I/239)
\bibitem[\protect\citeauthoryear{Imai et al.}{1997}]{imai97} 
Imai, H., Sasao, T., Kameya, O., Miyoshi, M., Shibata, K. M., Asaki, Y., 
Omodaka, T., Morimoto, M., Mochizuki, N., Suzuyama, T., Iguchi, S., Kameno, 
S., Jike, T., Iwadate, K., Sakai, S., Miyaji, T., Kawaguchi, N., Miyazawa, K. 
1997, A\&A, 317, 67
\bibitem[\protect\citeauthoryear{Imai et al.}{2008}]{imai08} 
Imai, H., Fujii, T., Omodaka, T., Deguchi, S 2008, PASJ, 60, 55
\bibitem[\protect\citeauthoryear{Kemball et al.}{1995}]{kemball95} 
Kemball, A. J., Diamond, P. J., Cotton, W. D. 1995, A\&AS, 110, 383
\bibitem[\protect\citeauthoryear{Kim et al.}{2010}]{kim10} 
Kim, J., Cho, S.-H., Oh, C. S., Byun, D.-Y. 2010 ApJS, 188, 209
\bibitem[\protect\citeauthoryear{Kirrane}{1987}]{kirrane87} 
Kirrane T.-M., 1987, PhD thesis, University of Manchester
\bibitem[\protect\citeauthoryear{Leal-Ferreira et al.}{2012}]{lf12} 
Leal-Ferreira, M. L., Vlemmings, W. H. T., Diamond, P. J., Kemball, A., Amiri, 
N., Desmurs, J.-F. 2012, A\&A, 540, 42
\bibitem[\protect\citeauthoryear{Meixner et al.}{1999}]{meixner99} 
Meixner, M., Ueta, T., Dayal, A., Hora, J. L., Fazio, G., Hrivnak, B. J., 
Skinner, C. J., Hoffmann, W. F., Deutsch, L. K. 1999, ApJS, 122, 221
\bibitem[\protect\citeauthoryear{Migenes et al.}{1999}]{migenes99} 
Migenes, V., Horiuchi, S., Slysh, V. I., Val'tts, I. E., Golubev, V. V., 
Edwards, P. G., Fomalont, E. B., Okayasu, R., Diamond, P. J., Umemoto, T., 
Shibata, K. M., Inoue, M. 1999, ApJS, 123, 487
\bibitem[\protect\citeauthoryear{Monnier et al.}{2004}]{monnier04} 
Monnier, J. D., Millan-Gabet, R., Tuthill, P. G., Traub, W. A., Carleton, 
N. P., Coud\'e du Foresto, V., Danchi, W. C., Lacasse, M. G., Morel, S., 
Perrin, G., Porro, I. L., Schloerb, F. P., Townes, C. H. 2004, ApJ, 605, 436
\bibitem[\protect\citeauthoryear{Nordhaus et al.}{2007}]{nordhaus07} 
Nordhaus, J., Blackman, E. G., Frank, A. 2007, MNRAS, 376, 599
\bibitem[\protect\citeauthoryear{Nyman et al.}{1986}]{nyman86} 
Nyman, L.-A., Johansson, L. E. B., Booth, R. S. 1986, A\&A, 160, 352
\bibitem[\protect\citeauthoryear{P\'erez-S\'anchez et al.}{2011}]{perez11} 
P\'erez-S\'anchez, A. F., Vlemmings, W. H. T., Chapman, J. M. 2011, MNRAS, 
418, 1402	
\bibitem[\protect\citeauthoryear{Ragland et al.}{2006}]{ragland06} 
Ragland, S., Traub, W. A., Berger, J.-P., Danchi, W. C., Monnier, J. D., 
Willson, L. A., Carleton, N. P., Lacasse, M. G., Millan-Gabet, R., Pedretti, 
E.; Schloerb, F. P., Cotton, W. D., Townes, C. H., Brewer, M., Haguenauer, 
P., Kern, P., Labeye, P., Malbet, F., Malin, D., Pearlman, M., Perraut, K., 
Souccar, K., Wallace, G. 2006, ApJ, 652, 650
\bibitem[\protect\citeauthoryear{Reid et al.}{1979}]{reid79} 
Reid, M. J., Moran, J. M., Leach, R. W., Ball, J. A., Johnston, K. J., 
Spencer, J. H., Swenson, G. W. 1979, ApJ, 227, 89
\bibitem[\protect\citeauthoryear{Reid}{1990}]{reid90} 
Reid, M J. 1990, IAUS, 140, 21
\bibitem[\protect\citeauthoryear{Reid \& Menten}{2007}]{reid07} 
Reid, M. J., Menten, K. M. 2007, ApJ, 671, 2068
\bibitem[\protect\citeauthoryear{Richards et al.}{2011}]{richards11} 
Richards, A. M. S., Elitzur, M., Yates, J. A. 2011, A\&A, 525, 56
\bibitem[\protect\citeauthoryear{Richards et al.}{2012}]{richards12} 
Richards, A. M. S., Etoka, S., Gray, M. D., Lekht, E. E., Mendoza-Torres, 
J. E., Murakawa, K., Rudnitskij, G., and Yates, J. A. 2012, A\&A 546, 16
\bibitem[\protect\citeauthoryear{Rudnitski et al.}{2010}]{rudnitski10} 
Rudnitski, G. M., Pashchenko, M. I., Colom, P. 2010, ARep, 54, 400
\bibitem[\protect\citeauthoryear{Shintani et al.}{2008}]{shintani08}
Shintani, M., Imai, H., Ando, K., Nakashima, K., Hirota, T., Inomata, N., Kai, 
T., Kameno, S., Kijima, M., Kobayashi, H., Kuroki, M., Maeda, T., Maruyama, K., 
Matsumoto, N., Miyaji, T., Nagayama, T., Nagayoshi, R., Nakamura, K., Nakagawa, 
A., Namikawa, D., Omodaka, T., Oyama, T., Sakakibara, S., Shimizu, R., Sora, 
K., Tsushima, M., Ueda, K., Ueda, Y., Yamashita, K. 2008, PASJ, 60, 1077
\bibitem[\protect\citeauthoryear{Sudou et al.}{2002}]{sudou02}
Sudou, H., Omodaka, T., Imai, H., Sasao, T., Takaba, H., Nishio, M., Hasegawa, 
W., Nakajima, J. 2002, PASJ, 54, 757
\bibitem[\protect\citeauthoryear{Surcis et al.}{2011}]{surcis11} 
Surcis, G., Vlemmings, W. H. T., Curiel, S., Hutawarakorn Kramer, B., 
Torrelles, J. M., Sarma, A. P. 2011, A\&A, 527, 48
\bibitem[\protect\citeauthoryear{Szymczak et al.}{2001}]{szymczak01} 
Szymczak, M., B\l aszkiewicz, L., Etoka, S., Le Squeren, A. M. 2001, 
A\&A, 379, 884
\bibitem[\protect\citeauthoryear{Vlemmings et al.}{2001}]{vlemmings01} 
Vlemmings, W., Diamond, P. J., van Langevelde, H. J. 2001, A\&A, 375, 1
\bibitem[\protect\citeauthoryear{Vlemmings et al.}{2002}]{vlemmings02} 
Vlemmings, W. H. T., Diamond, P. J., van Langevelde, H. J. 2002, A\&A,
394, 589
\bibitem[\protect\citeauthoryear{Vlemmings et al.}{2005}]{vlemmings05}
  Vlemmings, W.~H.~T., van Langevelde, H.~J., \& Diamond, P.~J.\ 2005,
  A\&A, 434, 1029
\bibitem[\protect\citeauthoryear{Vlemmings et al.}{2006}]{vlemmings06} 
Vlemmings, W. H. T., Diamond, P. J., Imai, H. 2006, Natur, 440, 58
\bibitem[\protect\citeauthoryear{Vlemmings et al.}{2012}]{vlemmings12} 
Vlemmings, W. H. T., Ramstedt, S., Rao, R., Maercker, M. 2012, A\&A, 540, 3	
\bibitem[\protect\citeauthoryear{Yates}{1993}]{yates93} 
Yates, J. A., 1993, PhD thesis, University of Manchester
\bibitem[\protect\citeauthoryear{Yates \& Cohen}{1994}]{yates94} 
Yates, J. A., Cohen, R. J. 1994, MNRAS, 270, 958
\bibitem[\protect\citeauthoryear{Wardle \& Kronberg}{1974}]{wardle74} 
Wardle, J. F. C., Kronberg, P. P. 1974, ApJ, 194, 249
\bibitem[\protect\citeauthoryear{Wolak et al.}{2012}]{wolak12} 
Wolak, P., Szymczak, M., G\'erard, E. 2012, A\&A, 537, 5
\bibitem[\protect\citeauthoryear{Zeeman}{1897}]{zeeman97} 
Zeeman, P. 1897, Philosophical Mag., 43, 226

\end{thebibliography}
\end{document}